\documentclass[twocolumn]{aastex63}
\usepackage{multirow}
\usepackage[figuresright]{rotating}
\usepackage{subfigure}
\usepackage{epic,eepic}
\usepackage{graphicx}
\usepackage{longtable}
\usepackage{float}
\usepackage{lineno}
\usepackage{pifont}
\usepackage{gensymb}
\usepackage{graphicx}
\usepackage{threeparttable}
\usepackage{textcomp, gensymb}
\usepackage{amsmath}
\usepackage{color}
\usepackage{hyperref}
\newcommand{\uat}[2]{\href{http://astrothesaurus.org/uat/#2}{#1 (#2)}}

\shorttitle{\rm UV photometry and habitability}
\shortauthors{Li et al.}

\begin{document}

\title{\rm Ultraviolet Photometry and Habitable Zones of Over 2700 Planet-Hosting Stars}

\correspondingauthor{Song Wang}
\email{songw@bao.ac.cn}

\author{Xue Li}
\affiliation{Key Laboratory of Optical Astronomy, National Astronomical Observatories, Chinese Academy of Sciences, Beijing 100101, China}
\affiliation{School of Astronomy and Space Sciences, University of Chinese Academy of Sciences, Beijing 100049, China}

\author{Song Wang}
\affiliation{Key Laboratory of Optical Astronomy, National Astronomical Observatories, Chinese Academy of Sciences, Beijing 100101, China}
\affiliation{Institute for Frontiers in Astronomy and Astrophysics, Beijing Normal University, Beijing 102206, China}

\author{Henggeng Han}
\affiliation{Key Laboratory of Optical Astronomy, National Astronomical Observatories, Chinese Academy of Sciences, Beijing 100101, China}

\author{Jifeng Liu}
\affiliation{Key Laboratory of Optical Astronomy, National Astronomical Observatories, Chinese Academy of Sciences, Beijing 100101, China}
\affiliation{School of Astronomy and Space Sciences, University of Chinese Academy of Sciences, Beijing 100049, China}
\affiliation{Institute for Frontiers in Astronomy and Astrophysics, Beijing Normal University, Beijing 102206, China}
\affiliation{New Cornerstone Science Laboratory, National Astronomical Observatories, Chinese Academy of Sciences, Beijing 100012, People's Republic of China}

\let\cleardoublepage\clearpage
\begin{abstract}

The ongoing discovery of exoplanets has sparked significant interest in finding suitable worlds that could potentially support life. 
Stellar ultraviolet (UV; 100--3000 \AA) radiation may play a crucial role in determining the habitability of their planets.
In this paper, we conducted a detailed analysis of the UV photometry of over 2700 host stars with confirmed planets, using observational data from the GALEX and Swift UVOT missions. 
We performed aperture photometry on single-exposure images, and provided photometric catalogs that can be used to explore a wide range of scientific questions, such as stellar UV activity and planet habitability.
By calculating the circumstellar habitable zone (CHZ) and UV habitable zone (UHZ), we found that fewer than 100 exoplanets fall within both of these zones, with the majority being gas giants.  
We also examined stellar activity based on their far-UV (FUV) and near-UV (NUV) emissions.
We found the FUV$-$NUV color more effectively represents stellar activity compared to the $R^{\prime}_{\rm FUV}$ and $R^{\prime}_{\rm NUV}$ indices. 
The Sun's low FUV emission and moderate NUV emission highlight its uniqueness among (solar-like) stars.

\end{abstract}


\section{INTRODUCTION}
\label{intro.sec}

The study of exoplanet habitability has rapidly advanced over the past few decades, driven by the discovery of thousands of exoplanets. The habitability of exoplanets involves various factors, such as their location in the habitable zone (HZ), atmospheric composition, and actual presence of liquid water on their surfaces \citep{2013Sci...340..577S, 2014Sci...344..277Q}.

The concept of HZ is central to assessing planetary habitability. Generally, it defines the region around a star where conditions may allow for the presence of liquid water on a planet's surface \citep{1993Icar..101..108K}. Previous studies have refined the boundaries of the circumstellar habitable zone (CHZ) based on stellar properties and updated climate models  \citep{2013ApJ...765..131K, 2014ApJ...787L..29K}. The atmospheric composition of an exoplanet also plays a crucial role in its potential habitability. For example, gaseous envelopes rich in greenhouse gases such as CO$_2$ can trap heat at the outer edge of the HZ and maintain surface temperatures suitable for liquid water \citep{2015ApJ...806..180W}. However, the presence of a thick atmosphere might lead to a runaway greenhouse effect, making the planet uninhabitable \citep{1993Icar..101..108K}.

UV radiation also significantly influences planetary habitability. 
Generally, UV radiation can be divided into X-ran UV (XUV; 1--100 \AA), extreme UV (EUV; 100--1000 \AA), far-UV (FUV; $\sim$ 1000--2000 \AA), and near-UV (NUV; $\sim$ 2000--3000 \AA).
On one hand, high XUV and EUV exposure can strip away planetary atmospheres and inhibit the development of life \citep{1977OrLi....8..259T, 2007AsBio...7..185L, 2012NatCh...4..895R, 2015NatCh...7..301P}.
On the other hand, biologically, UV radiation can break molecule bonds, leading to the degradation of essential compounds such as nucleic acids and proteins. Therefore, high levels of UV radiation can prevent the formation of complex organic molecules necessary for life. However, moderate levels of NUV radiation can drive the synthesis of prebiotic compounds, especially for ribonucleic acid, which is the building blocks for the emergence of life, as reported by some experimental studies \citep[e.g.,][]{1977OrLi....8..259T, 2012NatCh...4..895R, 2015NatCh...7..301P, 2016AsBio..16...68R}.
The regions around stars with favourable conditions for NUV irradiation are defined as the UV habitable zone \citep[UHZ;][]{2023MNRAS.522.1411S}.
In the following parts of this paper, "UV" refers specifically to the FUV and NUV bands.

Stellar UV radiation are mostly from stellar chromosphere and photosphere, with radiation levels varying according to stellar types and their evolutionary stages.
For example, the UV radiation of M stars is predominantly generated by their chromosphere due to magnetic activity, whereas, for earlier-type stars, UV radiation is dominated by their photospheres.
M-dwarf stars, which are typically magnetically active, often exhibit high levels of UV radiation and frequent UV flares, potentially affecting planetary atmospheres and surface conditions for habitability \citep{2014AJ....148...64S, 2018AJ....155..122S, 2020ApJ...895....5P, 2021ApJ...907...91L}.
A comprehensive study of UV radiation for stars hosting exoplanets is essential for assessing their UV habitability.

In this work, we analyzed UV photometry of stars with confirmed planets using data from the Galaxy Evolution Explore (GALEX) and Swift Ultraviolet/Optical Telescope (UVOT), and discussed the impact of UV radiation on planetary habitability. In Section \ref{samples.sec}, we introduce the catalog of host stars with confirmed planets. In Section \ref{galex.sec} and Section \ref{swift.sec}, we detail the photometric analysis processes of the GALEX and Swift UVOT data. We then calculate the CHZ and UHZ and discuss the habitability of our sample stars in Section \ref{hz.sec}. In Section \ref{activity.sec}, we study stellar activities for our sample sources.

\section{Sample selection}
\label{samples.sec}

\begin{figure}
    \centering
    \subfigure{
    \includegraphics[width=0.46\textwidth]{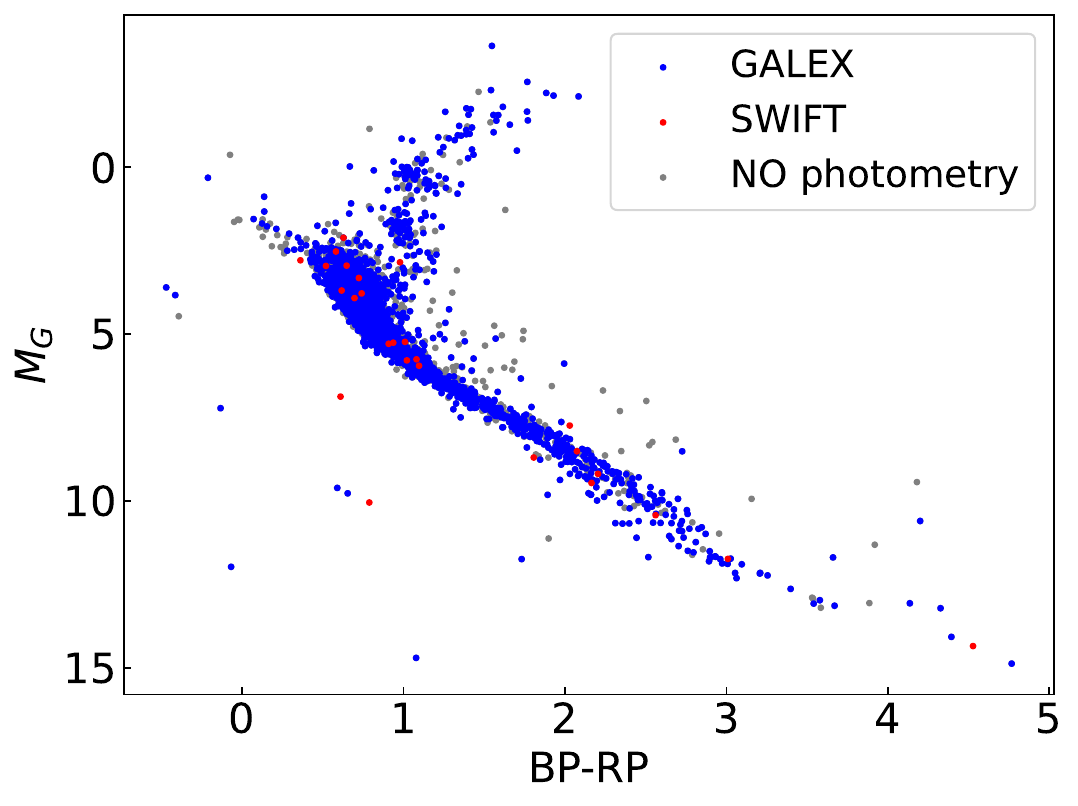}
    \label{hr_gaia.fig}}
    \subfigure{
    \includegraphics[width=0.46\textwidth]{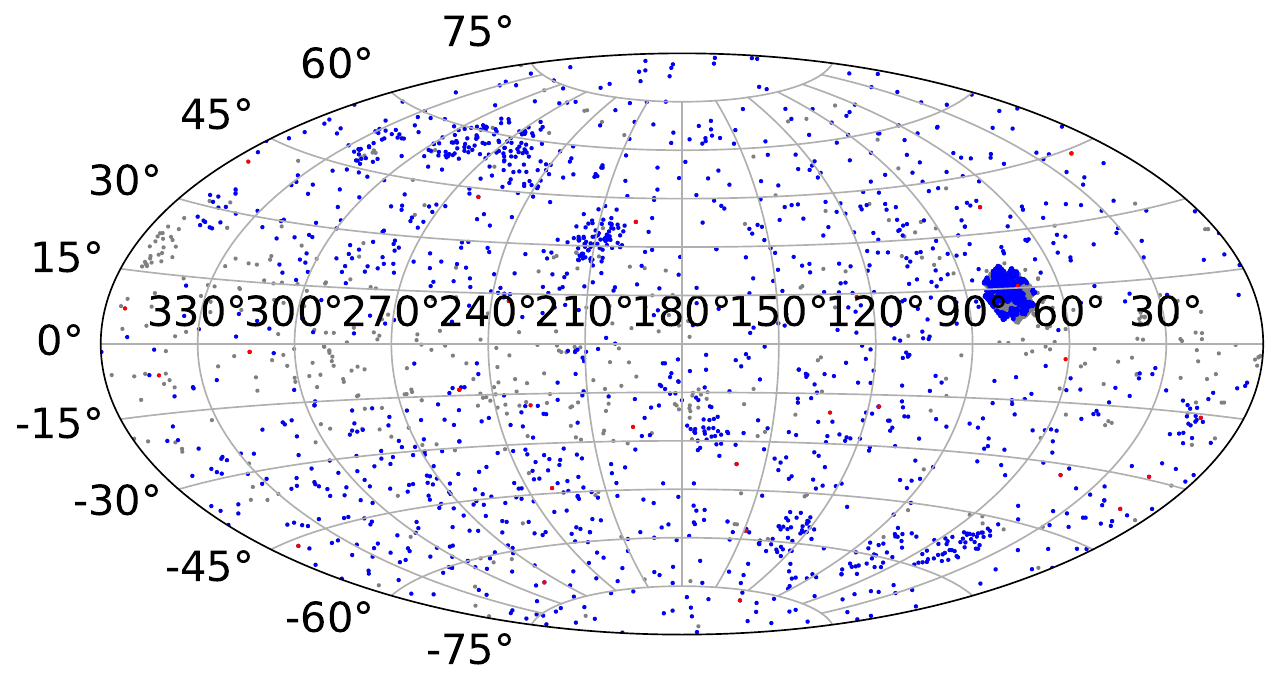}
    \label{galatic.fig}}
    \caption{Top panel: Hertzsprung–Russell diagram of the sample stars hosting planets from \cite{ps}. The blue and red points represent stars with GALEX and SWIFT UVOT observations, respectively, while the gray points represent stars without UV photometry. 
    Bottom panel: Galactic distribution of the sample stars. }
    \label{hr.fig}
\end{figure}

The NASA Exoplanet Archive catalog\footnote{\url{https://exoplanetarchive.ipac.caltech.edu/}} provides all the planets discovered so far through various methods (e.g., transit, radial velocity, microlensing, direct imaging). This publicly accessible database is maintained by the NASA Exoplanet Science Institute and is updated in real time.
As of November 2023, the catalog includes 5,535 confirmed exoplanets orbiting around 3,923 host stars, which will be used as our initial sample.
The stellar sample covers a broad range of spectral types, from O stars to M stars and brown dwarfs, with effective temperatures ($T_{\rm eff}$) ranging from approximately 57,000 K to 400 K. The sample includes 169 giants and 2,563 dwarfs.
Figure \ref{hr.fig} shows the Hertzsprung–Russell diagram and the galactic distributions of the sample.

\section{GALEX photometry}
\label{galex.sec}

The GALEX mission, a NASA space telescope, was designed to observe the universe in the UV bands \citep{2007ApJS..173..682M}. It operates in two UV bands: far ultraviolet (FUV, $\lambda_{\rm eff}\sim1528\AA$, 1344--1786\AA) and near ultraviolet (NUV, $\lambda_{\rm eff}\sim2310\AA$, 1771--2831\AA) bands. The latest catalog, GR6$+$7 \citep{2017ApJS..230...24B}, includes a total of 82,992,086 objects, with observations from the All-Sky Imaging Survey (AIS, $t_{\rm exp} =$ 100 s), Medium-depth Imaging Survey (MIS, $t_{\rm exp} =$ 1500 s), and Deep-depth Imaging Survey (DIS, $t_{\rm exp} =$ 10,000 s).

\subsection{Data Preparation}
\label{galex_img.sec}

In our sample, many nearby stars exhibit high proper motion, causing their positions to vary significantly between exposures. Therefore, we downloaded single-exposure images, rather than co-added images, to perform photometry.

The GALEX $VisitPhotoObjAll$ catalog \citep{2017ApJS..230...24B} contains FUV and NUV photometric data for 292,296,119 detections from individual exposures observed by the AIS and MIS. The $VisitPhotoExtract$ catalog provides detailed information for each GALEX observation, including observation date, exposure time, image download path, etc. 

First, we cross-matched the initial sample (see Section \ref{samples.sec}) with the GALEX $VisitPhotoObjAll$ catalog using a match radius of 3$\arcsec$ via the CasJobs\footnote{\url{https://galex.stsci.edu/casjobs/}} and obtained the keyword ``photoExtractID." Second, we used ``photoExtractID" to retrieve the image download paths of each observation from the $VisitphotoExtract$ catalog, including the ``intensity" and sky background maps of both FUV and NUV bands. Third, we used \texttt{wget} to download the images for each observation. Finally, 4873 NUV images and 3144 FUV images of 2717 stars were downloaded.

Some targets were not recorded by CasJobs but can be found using \textit{gPhoton}, a software that allows for the analysis of GALEX UV data at the photon level \citep{2016ApJ...833..292M}.
As a supplement, we used \textit{gPhoton} to download intensity images in both NUV and FUV bands for these sources. We downloaded single-exposure images, each with a size of 0.1 degrees $\times$ 0.1 degrees, centered on the position of each target. 
In total, we obtained multiple observations of FUV and NUV bands of 2742 host stars. All the GALEX data used in this paper can be found in MAST \citep{https://doi.org/10.17909/t9h59d, 2016ApJ...833..292M}.

\subsection{Photometry steps}
\label{galex_pho.sec}

First, we used proper motion data from the Gaia DR3 catalog to calculate the coordinates of each target at the time of observation, which is crucial for stars with high proper motion. Next, we used the Python package \texttt{photutils} to perform aperture photometry for our sample sources.

\begin{figure*}[htp]
    \centering
    \includegraphics[width=1\linewidth]{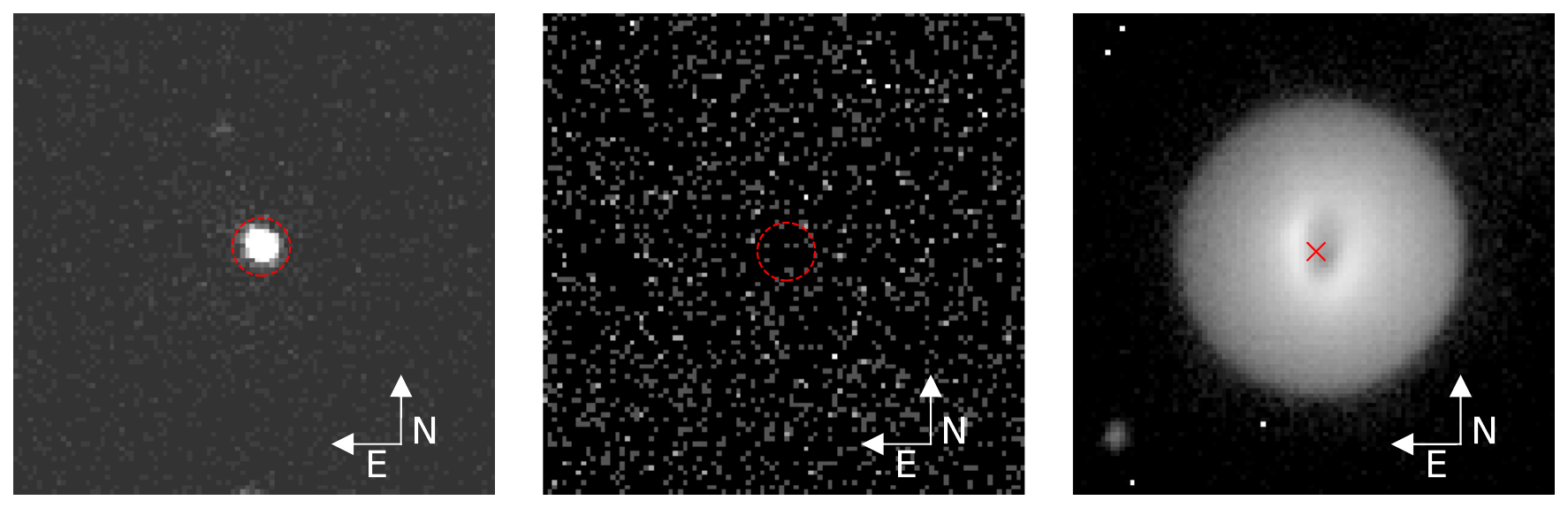}
    \caption{Examples of images for a general source (AU Mic in the left panel), an undetected source with an upper limit magnitude estimation (TRAPPIST-1 in the middle panel), and a saturated source (HD 19994 in the right panel), respectively.}
    \label{images.fig}
\end{figure*}

In the photometric analysis, a key step is constructing the sky background map.
For observations downloaded from the GALEX website (i.e., 11,320 images of the NUV band and 5788 images of the FUV band), we also obtained their background maps using CasJobs.
For observations downloaded via \textit{gPhoton} (i.e., 644 images for the NUV band and 271 images for the FUV band), no background maps were provided. Therefore, we created their background maps using the interpolation method with the {\tt\it photutils.background.Background2D} module.
We selected a box size of $10 \times 10$ pixels and a filter size of $5 \times 5$ pixels to match the scale of our images. Given the faintness of our targets, we set the {\tt exclude\_percentiles} parameter for background estimation to 50\% for NUV band and 90\% for  FUV band. 
We employed {\tt\it astropy.stats.SigmaClip} for sigma clipping with a 3$\sigma$ threshold.
The {\tt bkg\_estimator} method was set to {\tt SExtractorBackground}.

\begin{figure*}[htp]
    \centering
    \subfigure[FUV band]{
    \includegraphics[width=0.47\textwidth]{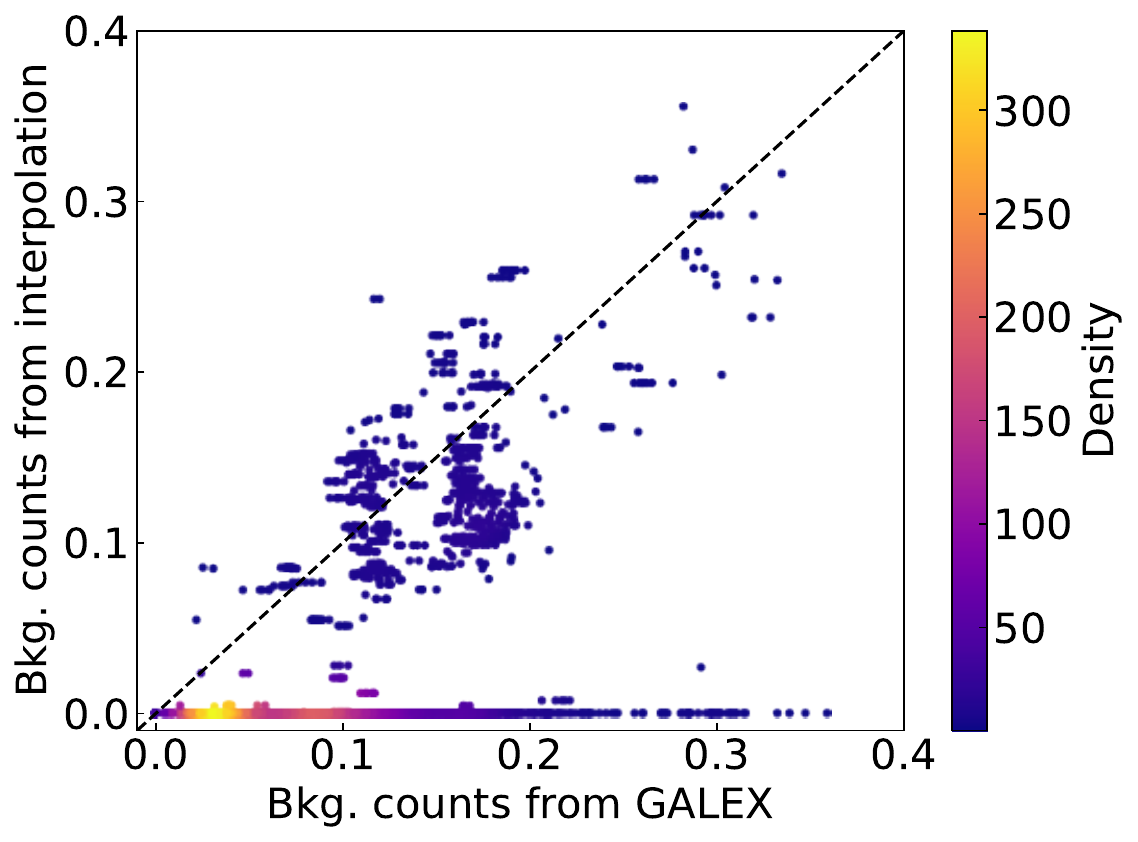}
    \label{fuv_bkg_compare.fig}}
    \subfigure[NUV band]{
    \includegraphics[width=0.46\textwidth]{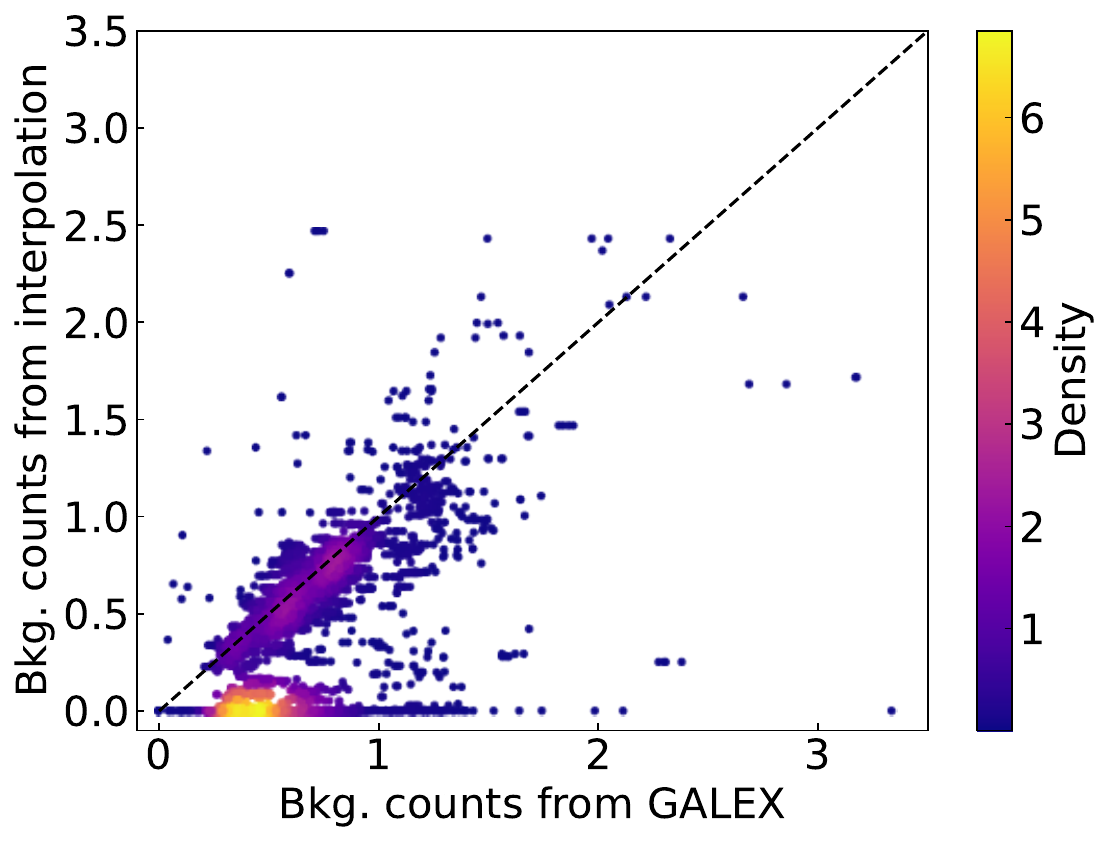}
    \label{nuv_bkg_compare.fig}}
    \subfigure[FUV band]{
    \includegraphics[width=0.46\textwidth]{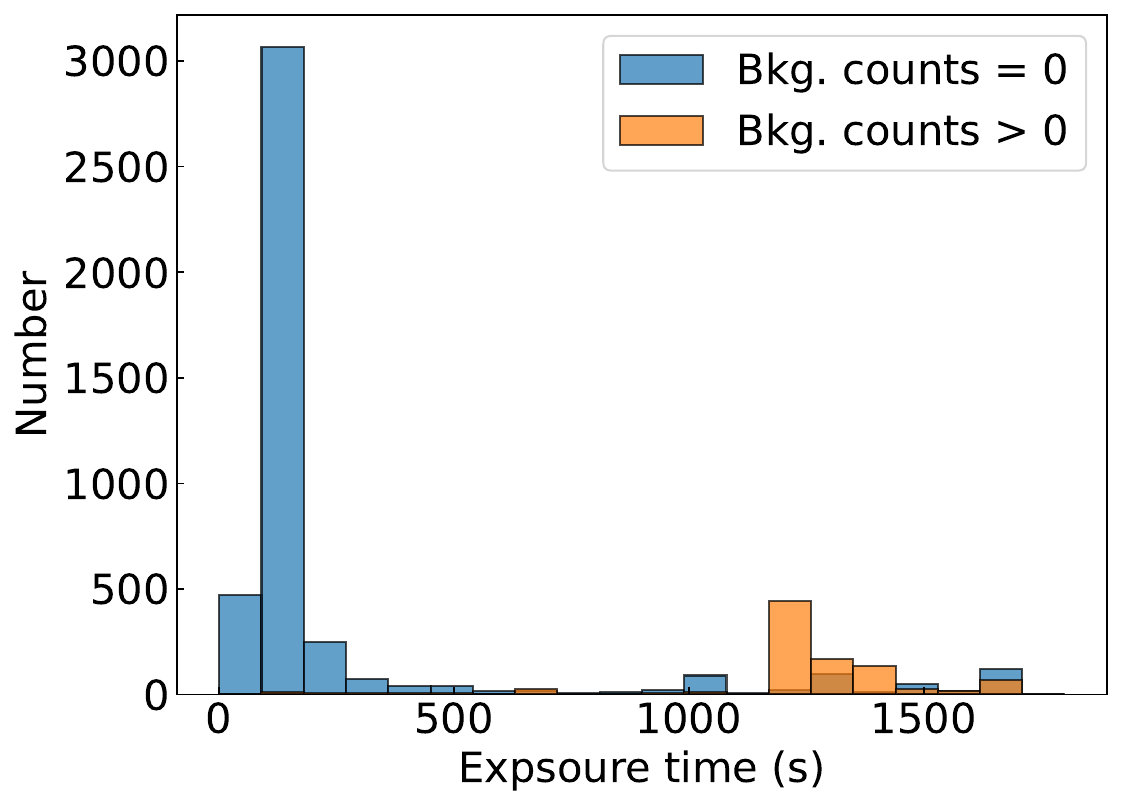}
    \label{fuv_exptime_hist.fig}}
    \subfigure[NUV band]{
    \includegraphics[width=0.46\textwidth]{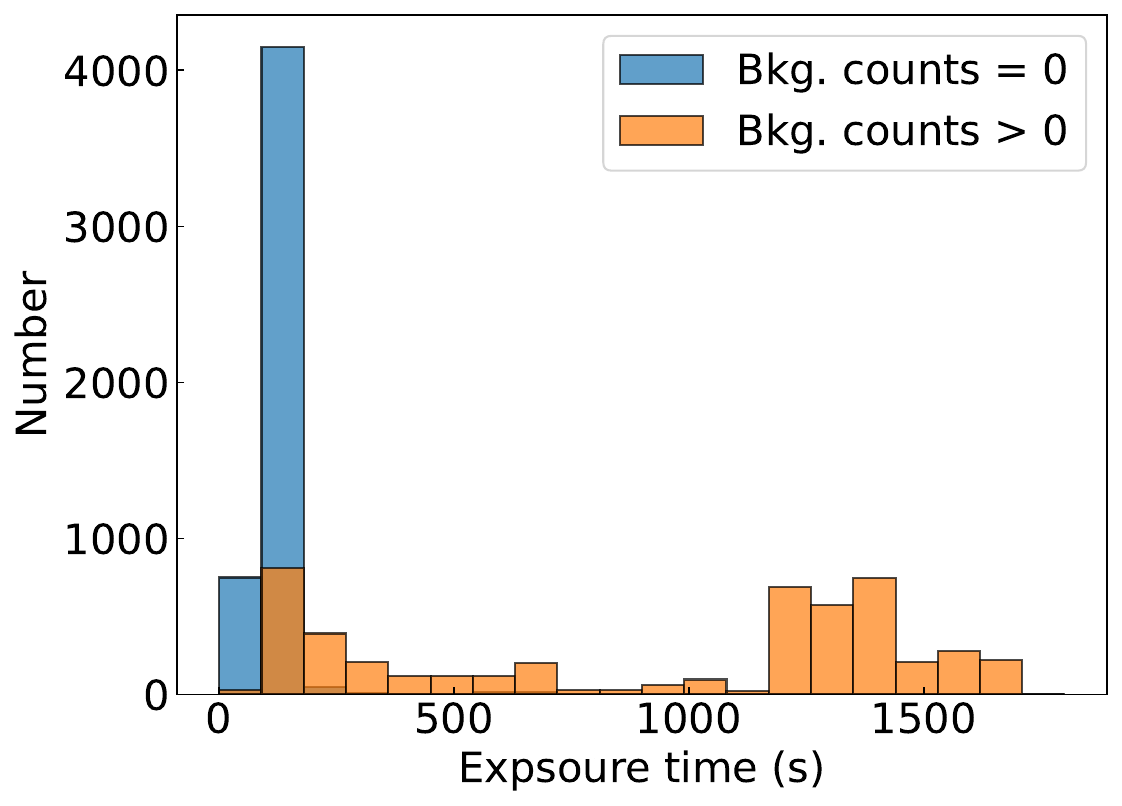}
    \label{nuv_exptime_hist.fig}}
    \caption{Panel a: Comparison between background counts obtained from the interpolation method and those derived from GALEX background images in the FUV band. The aperture size for background estimation is 6 pixels and the unit is counts/s. Colors represent number density.}
    Panel b: Comparison between background counts obtained from the interpolation method and those derived from GALEX background images in the NUV band. 
    panel c: Distribution of exposure times for images with different background values using the interpolation method in the FUV band.
    Panel d: Distribution of exposure times for images with different background values using the interpolation method in the NUV band.
    \label{bkg.fig}
\end{figure*}

To verify the accuracy of our background maps, we also downloaded images with \textit{gPhoton} for those observations that already had backgrounds from GALEX and generated their background maps using the interpolation method. Figure \ref{bkg.fig} compares the interpolated backgrounds with those downloaded from GALEX, both calculated with an aperture size of 6 pixels around each sample source. The results generally show good agreement. However, a number of observations exhibit background values of zero when using the interpolation method. By checking the exposure times for each observation, we found that those with shorter exposure times are more likely to have background values of zero. 
As described above, we created background maps for observations downloaded via \textit{gPhoton}, covering both the NUV and FUV bands.
Among these observations, 160 NUV and 161 FUV background estimates have values of zero. The luminosities of the sample sources in these observations may therefore be slightly overestimated.

We examined the GALEX images and classified our targets into three types: general sources with good photometry (e.g., AU Mic, as shown in the left panel of Figure \ref{images.fig}); 
sources near detection limits (e.g., TRAPPIST-1, as shown in the middle panel of Figure \ref{images.fig});
saturated sources (e.g., HD 19994, as shown in the right panel of Figure \ref{images.fig}); . In the following sections, we will perform photometry separately for each type of sources.

\subsubsection{General Sources}
\label{general.sec}

We defined 22 circular apertures with radii ranging from 1 to 22 pixels in steps of 1 pixel, according to the GALEX documentation\footnote{\label{galex_doc}\url{http://www.galex.caltech.edu/researcher/techdoc-ch5.html}}. Using the \texttt{photutils} package, we performed aperture photometry for each target with all 23 aperture sizes. \texttt{Photutils} sums the pixel values within each aperture on background-subtracted images to measure the total counts per second (CPS). Then, we calculated the fluxes and magnitudes following\footnote{\url{https://asd.gsfc.nasa.gov/archive/galex/FAQ/counts_background.html}}

\begin{align}
\label{cps_flux.eq}
    \text{flux} \, (\text{erg/cm}^2/\text{s}/\text{\AA}) &= C1 \times \text{CPS}
\end{align}
and
\begin{align}
\label{cps_mag.eq}
    m_{\text{AB}} &= -2.5 \times \log_{10}(\text{CPS}) + C2.
\end{align}
Here, $C1$ represents the factor converting counts to flux, and $C2$ represents the zero point of the AB magnitude. For FUV band,  $C1$ is $1.4 \times 10^{-15}$ and $C2$ is 18.82. For NUV band,  $C1$ is $2.06 \times 10^{-16}$ and $C2$ is 20.08.

\begin{figure*}[htp]
    \centering
    \subfigure[]{
    \includegraphics[width=0.48\textwidth]{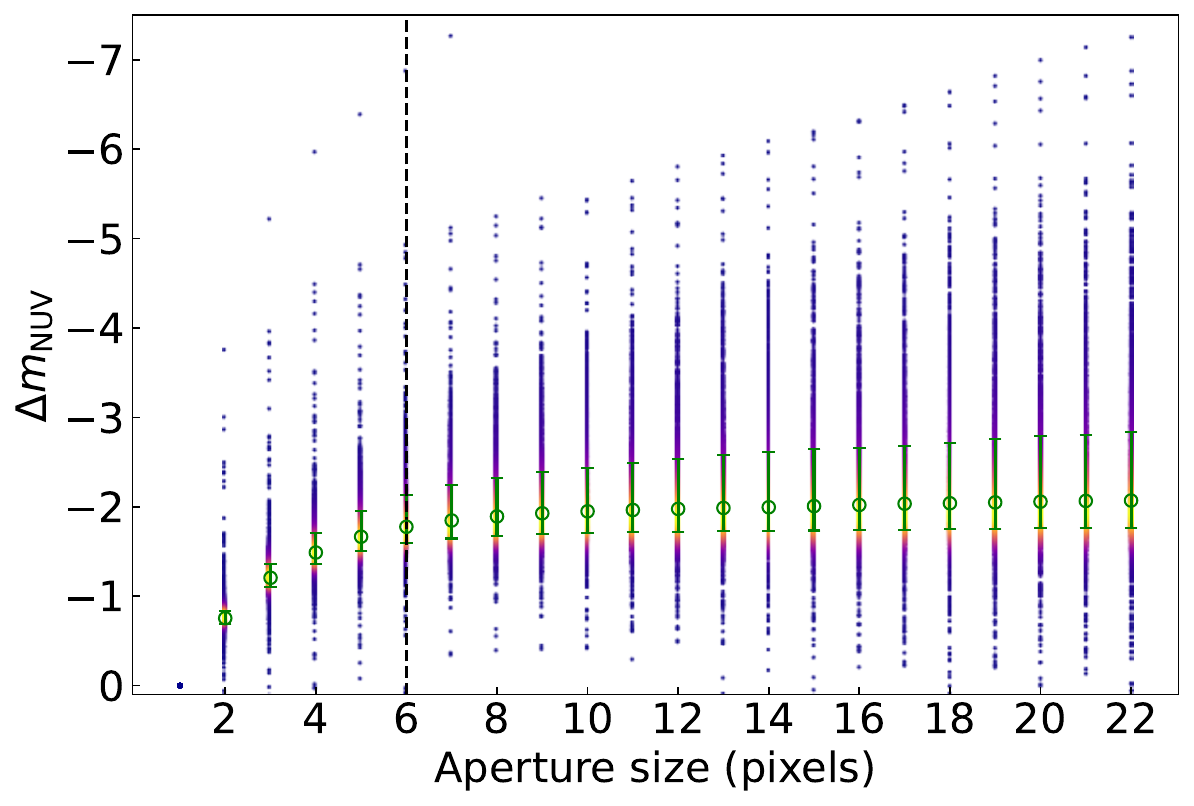}
    \label{nuv_growth_curve.fig}}
    \subfigure[]{
    \includegraphics[width=0.48\textwidth]{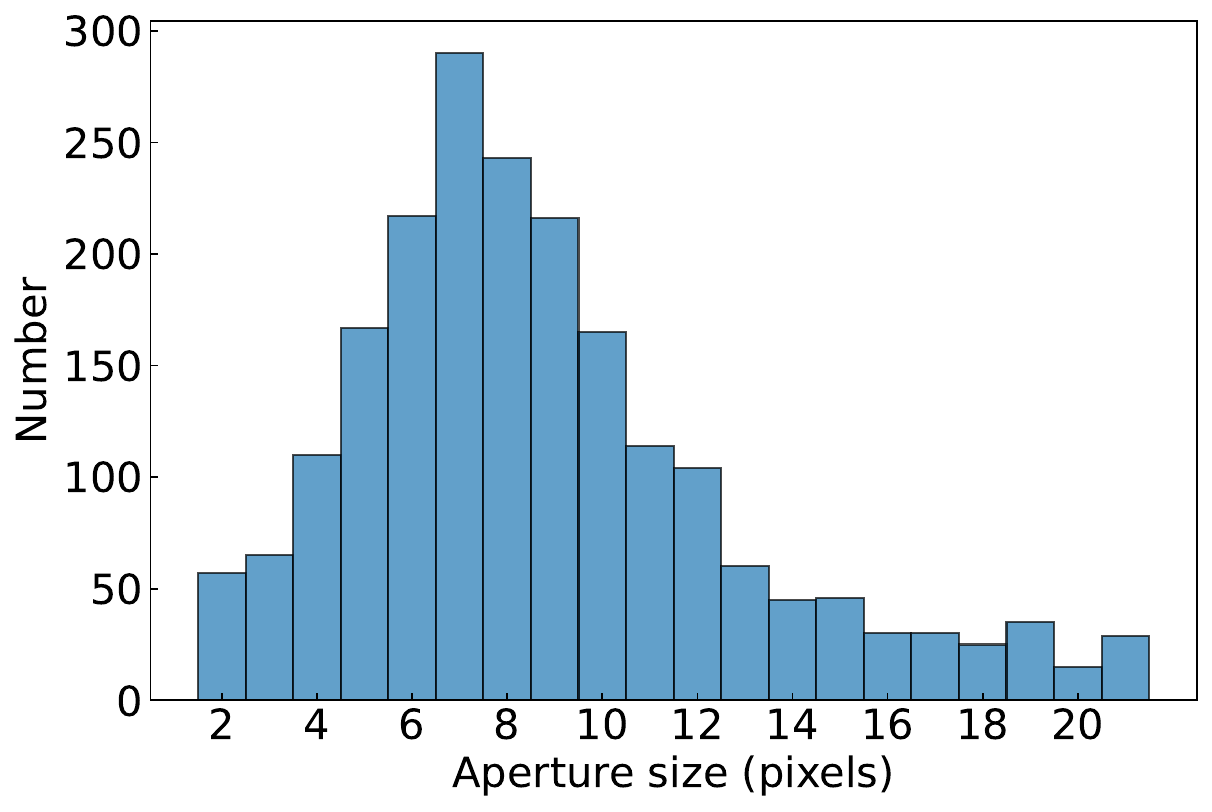}
    \label{nuv_knee_hist.fig}}
    \caption{Left Panel: Growth curves for all sources in the NUV band. The magnitude differences are derived by subtracting the photometry results using a 1-pixel aperture from those at larger apertures. Colors indicate number density. The green circles represent the median magnitude differences at each aperture.
    Right Panel: Distribution of the selected apertures from the growth curves for each target.}
    \label{growth_curve.fig}
\end{figure*}

\begin{figure*}[htp]
    \centering
    \subfigure[]{
    \includegraphics[width=0.46\textwidth]{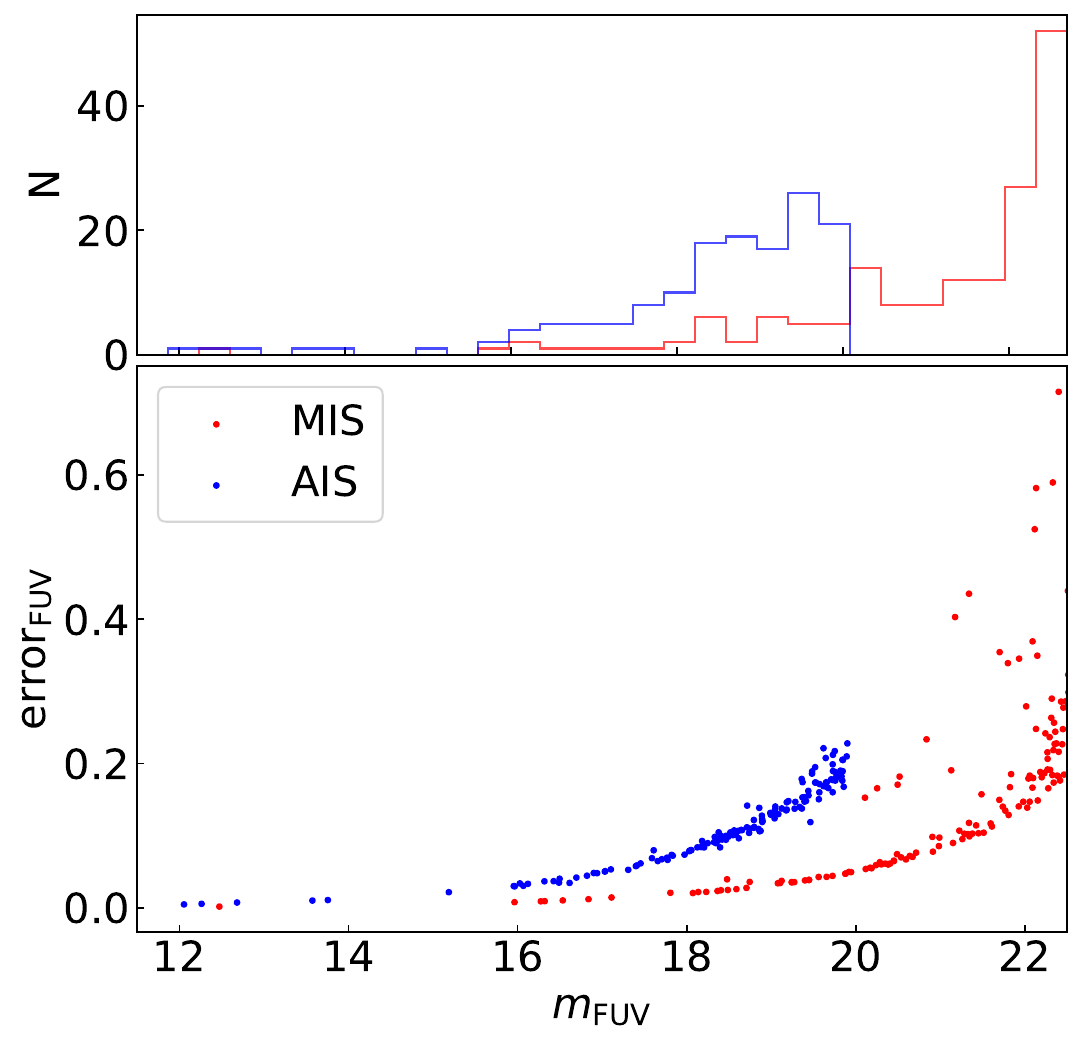}
    \label{fuv_err.fig}}
    \subfigure[]{
    \includegraphics[width=0.46\textwidth]{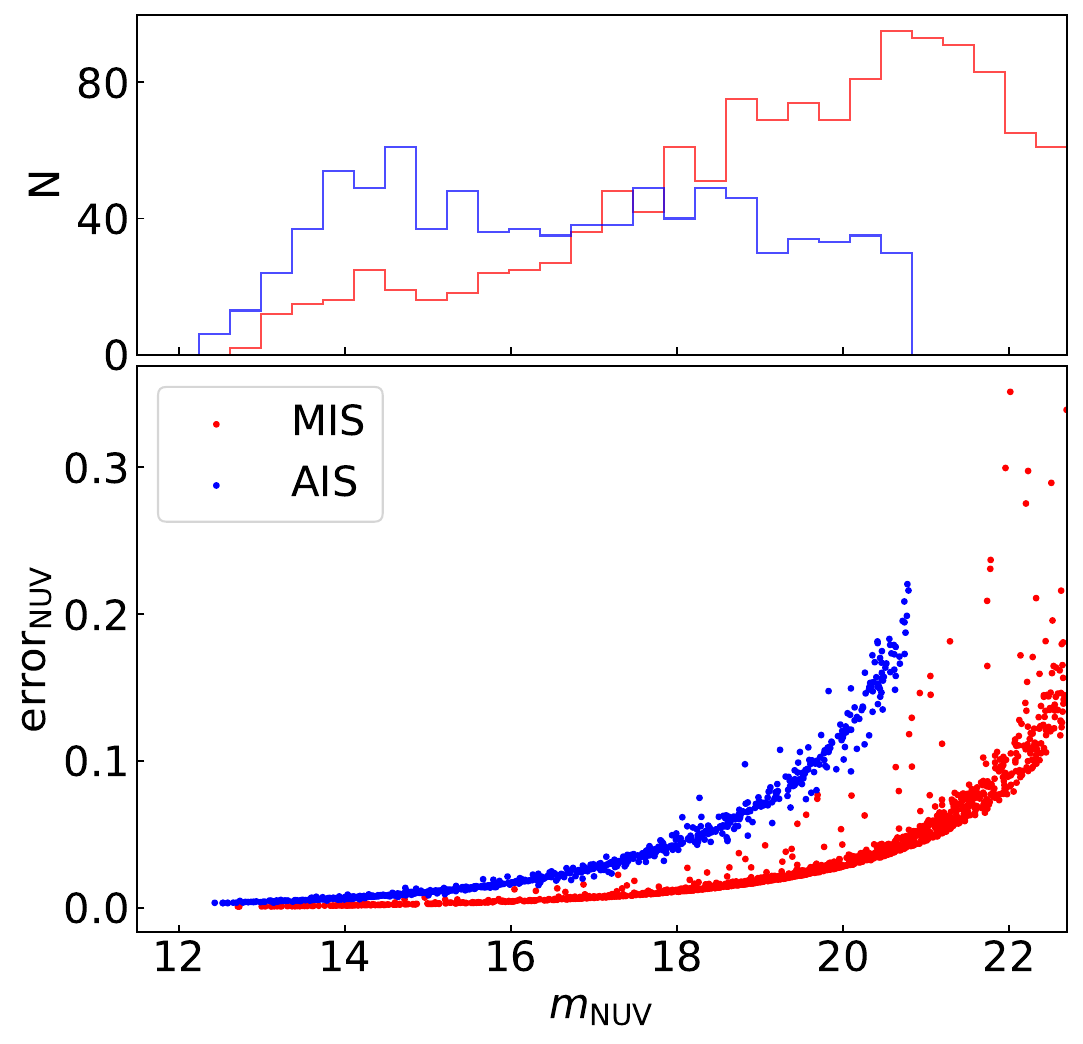}
    \label{nuv_err.fig}}
    \subfigure[]{
    \includegraphics[width=0.46\textwidth]{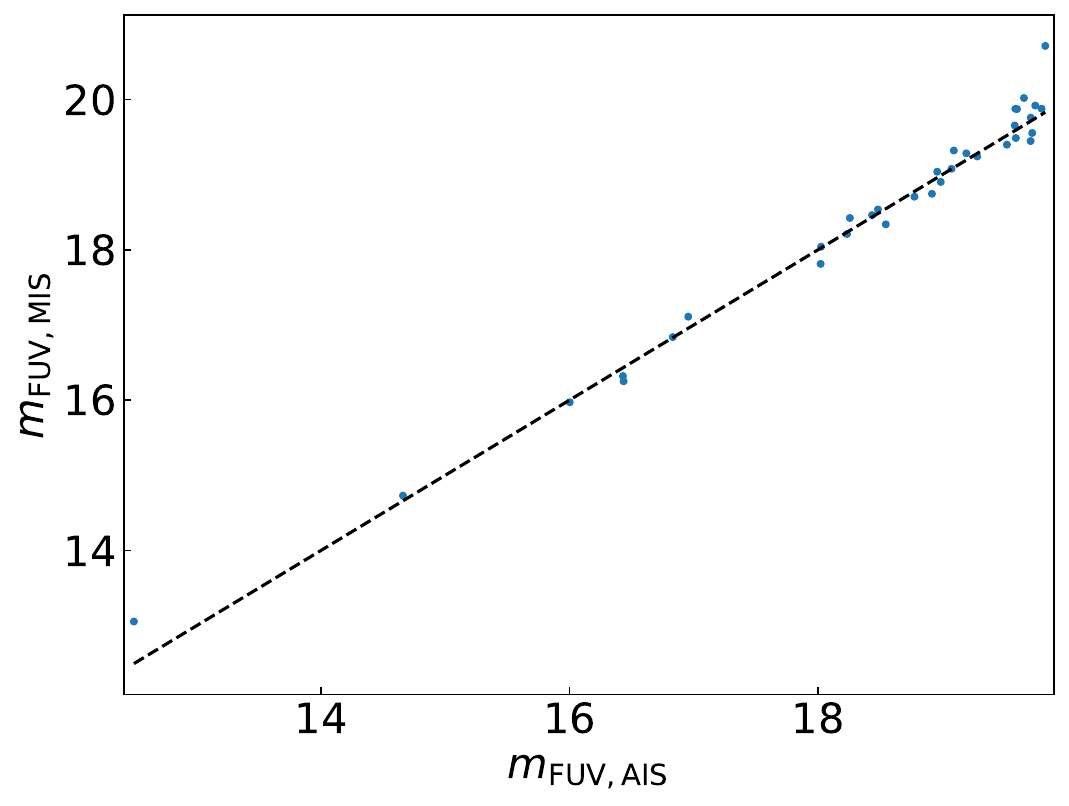}
    \label{ais_mis_fuv.fig}}
    \subfigure[]{
    \includegraphics[width=0.46\textwidth]{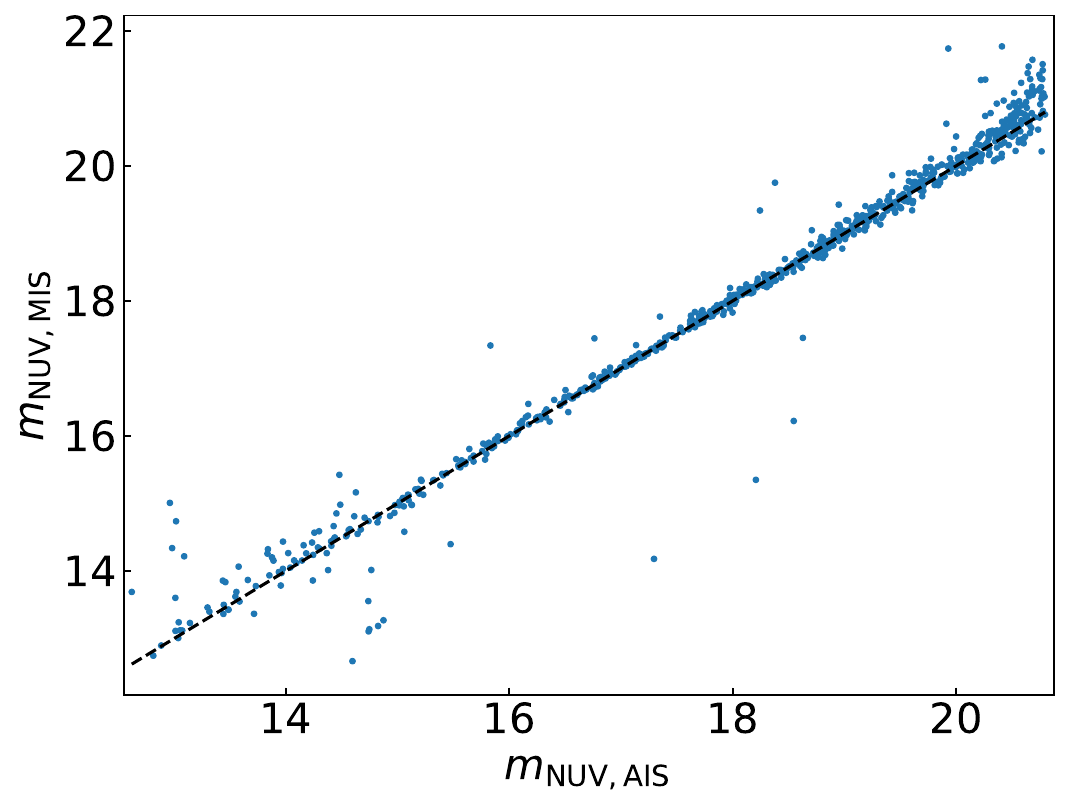}
    \label{ais_mis_nuv.fig}}
    \caption{
    Panel a: Distribution of uncertainty as a function of magnitude for the FUV band. The blue and red points represent the photometry results of AIS and MIS, respectively. The magnitudes are cut at the sensitivity limits for both AIS ($=$19.9 AB mag) and MIS ($=$22.7 AB mag).
    Panel b: Distribution of uncertainty as a function of magnitude for the NUV band. The magnitudes are cut at the sensitivity limits for both AIS ($=$20.7 AB mag) and MIS ($=$22.7 AB mag).
    Panel c: Comparison of FUV magnitudes measured from AIS and MIS.
    Panel d: Comparison of NUV magnitudes measured from AIS and MIS.}
    \label{err.fig}
\end{figure*}

Figure \ref{nuv_growth_curve.fig} shows the growth curves of the sources in the NUV band, with colors representing the number density. We determined the optimal aperture for each source by comparing the photometry results of adjacent apertures. When the magnitude difference between two adjacent apertures was smaller than the photometric error of the smaller aperture, that aperture was selected. From the distribution of the optimal apertures (as shown in Figure \ref{nuv_knee_hist.fig}), we found that the 7-pixel aperture was the most suitable for the NUV band. Considering that GALEX provides a Point Spread Function (PSF) correction for the 6-pixel aperture, we selected the photometry results using the 6-pixel aperture. This aperture is about double of the full width at half maximum of the PSF of NUV band ($\sim$ 4.9\arcsec $\sim$ 3.3 pixel)\footref{galex_doc}. We also used this aperture for FUV band, since the PSF in the FUV band ($\sim$4.2\arcsec $\sim$ 2.8 pixel) is similar to that in the NUV band. To sum up, we selected the photometry results obtained using a 6-pixel (9 arcsecs) aperture and added aperture corrections of 0.07 mag for the FUV band and 0.08 mag for the NUV band\footref{galex_doc}. 
The distribution of photometric errors with stellar brightness is shown in Figure \ref{err.fig}. It can be seen that the errors in MIS and AIS data vary differently with magnitude. The AIS data have much shorter exposure times and lower signal-to-noise ratios (S/Ns), resulting in larger errors for fainter objects.

The sensitivity limits for the GALEX surveys are approximately 19.9 AB mag in FUV and 20.8 AB mag in NUV for the AIS, and 22.7 AB mag in both bands for the MIS, as reported in previous studies \citep[e.g.,][]{2017ApJS..230...24B}. 
Therefore, we only kept sources brighter than these magnitudes as general sources.
Finally, for sources observed multiple times, we selected the observation with the longest exposure time as the final result.

\begin{figure*}[htp]
    \centering
    \subfigure[]{
    \includegraphics[width=0.48\textwidth]{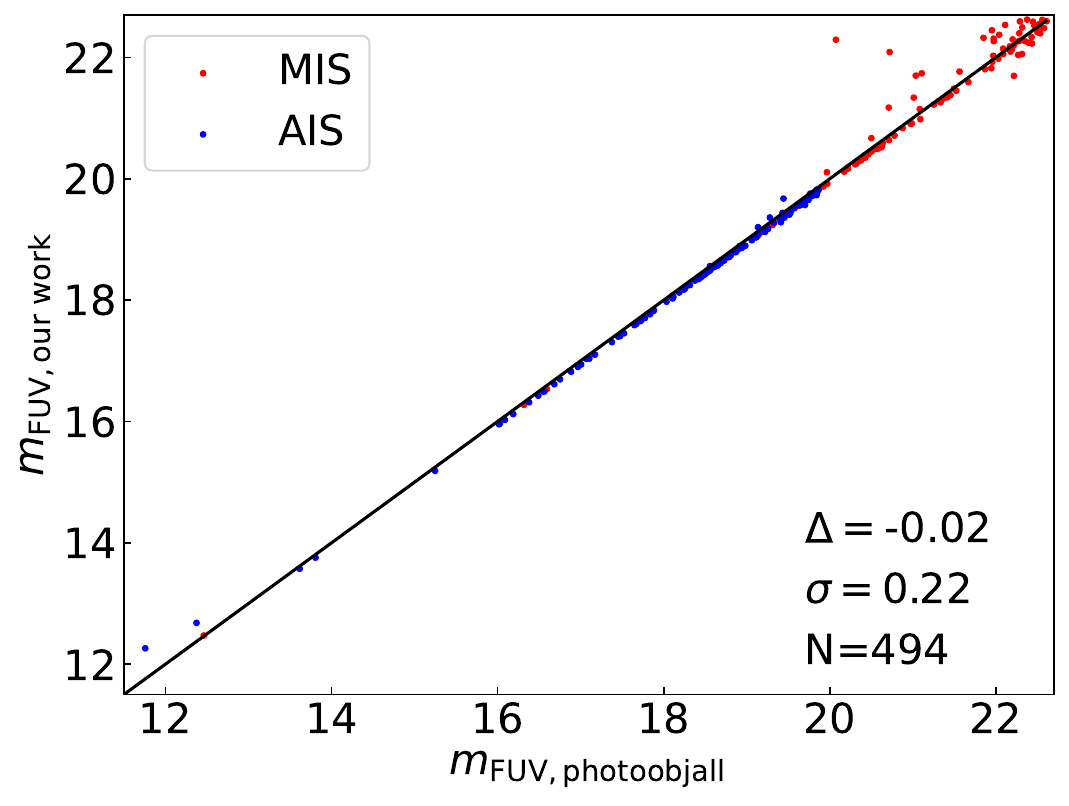}
    \label{fuv_compare.fig}}
    \subfigure[]{
    \includegraphics[width=0.48\textwidth]{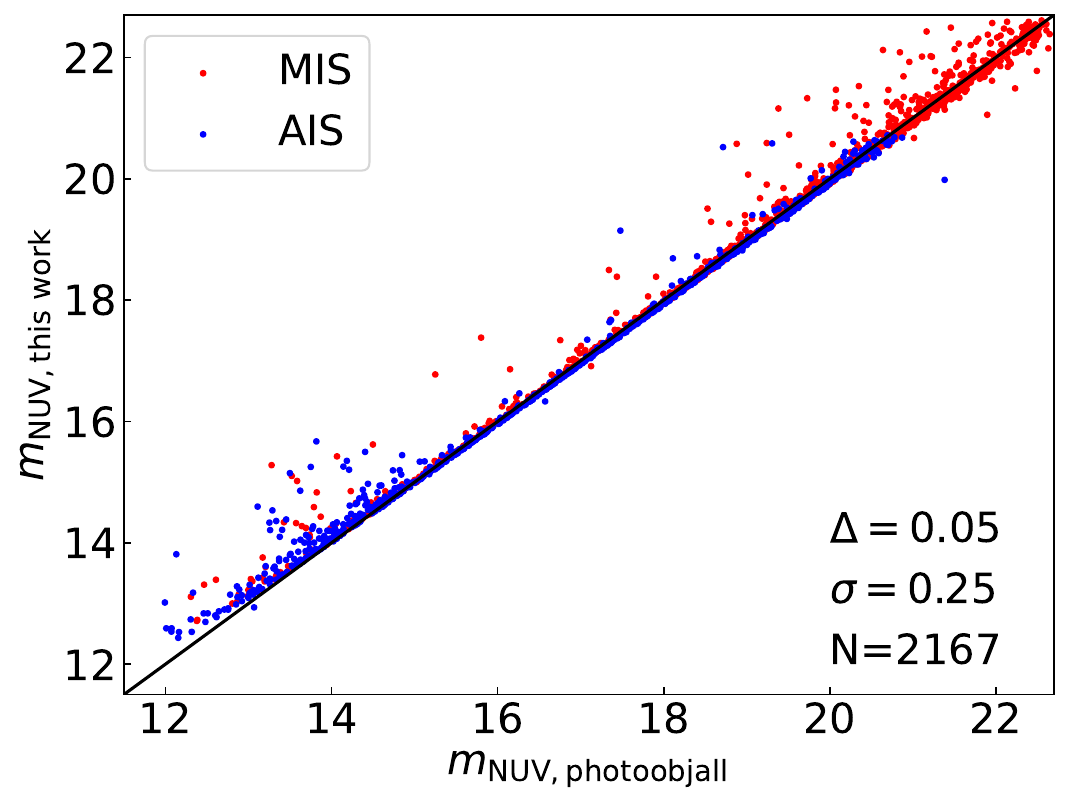}
    \label{nuv_compare.fig}}
    \caption{Left panel: Comparison of FUV magnitudes derived by this work and those from the GALEX $photoobjall$ catalog.
    Right panel: Comparison of NUV magnitudes derived by this work and those from the GALEX $photoobjall$ catalog.}
    \label{compare.fig}
\end{figure*}

Figure \ref{compare.fig} compares our photometric results with the single-exposure results provided by the GALEX $visitphotobjall$ catalog. The good agreement suggests our phtotometry results are accurate and reliable.
However, we noticed that in the NUV band, for sources brighter than $\approx$14 mag, our photometry appears slightly fainter than the values from the GALEX $photoobjall$ catalog. The GALEX photometry (i.e., the $nuv\_mag$ and $fuv\_mag$) are based on the ``AUTO" measurements from the SExtractor program, which measures the total flux of stars using an elliptical ``Kron" aperture \citep{1980ApJS...43..305K}. For bright, slightly diffuse stars, our circular aperture photometry may experience more photon loss compared to the "Kron" aperture photometry. Additionally, we examined the sources with large deviations and found that most of them have a large distance to the center of the field-of-view, often exceeding $\approx$0.5 degrees. For such sources, the reliability of GALEX photometry decreases.

Table \ref{galex.tab} presents the main photometric results (AB mag), using the longest exposure when multiple observations are available. 
Table \ref{galex_multi.tab} provides the photometric results for individual exposures.

\subsubsection{Saturated Sources}
\label{saturated.sec}

\begin{figure*}
    \centering
    \subfigure[]{
    \includegraphics[width=0.48\textwidth]{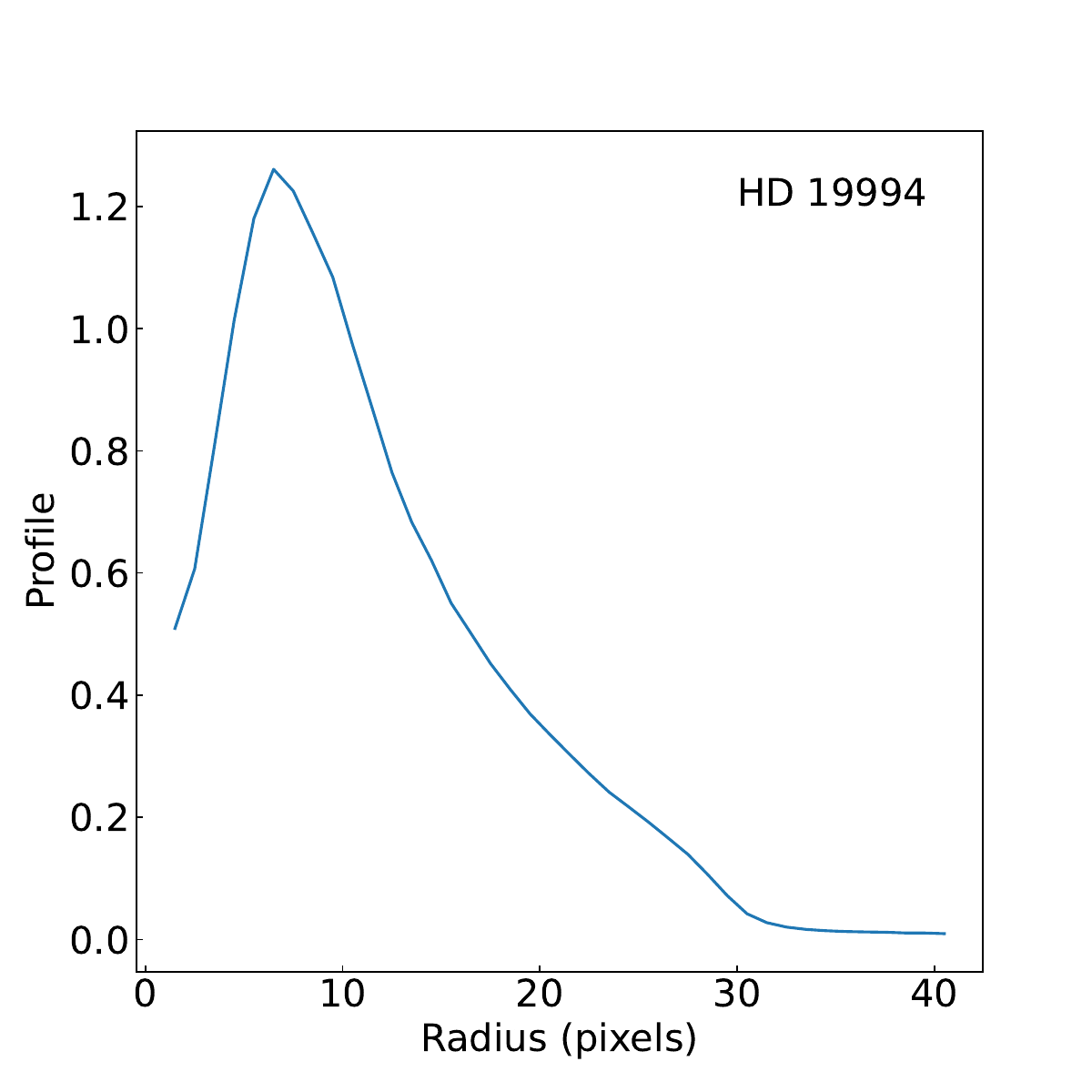}
    \label{profile.fig}}
    \centering
    \subfigure[]{
    \includegraphics[width=0.44\textwidth]{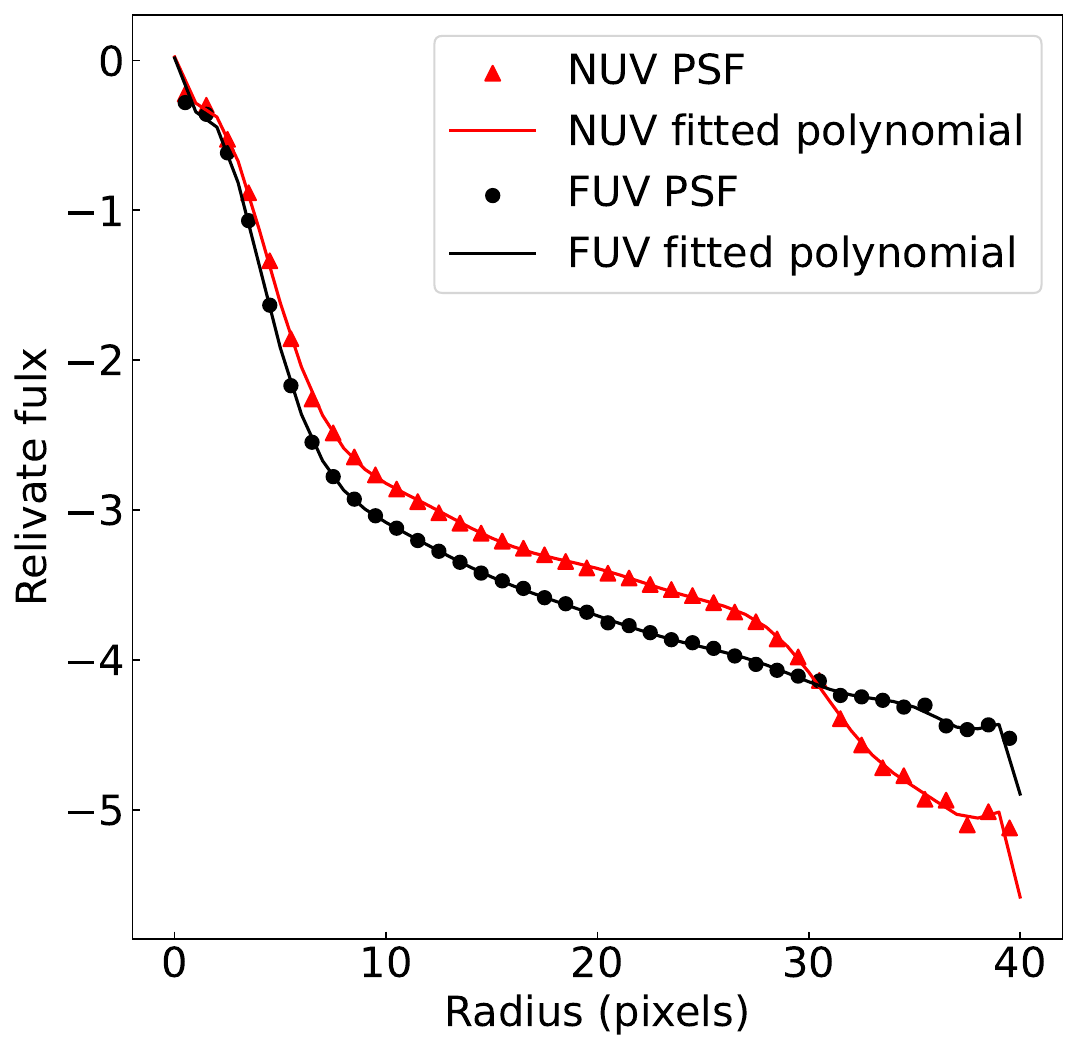}
    \label{psf_model.fig}}
    \caption{Left Panel: The surface brightness profile in the NUV band for a saturated source, HD 19994.
    Right Panel: The radial profiles of the FUV (black points) and NUV (red triangles) PSF models. The black and red lines represent the fifteenth-degree polynomial fittings to the radial profiles of the FUV and NUV PSF models.}
    \label{saturated.fig}
\end{figure*}

In our sample, some sources were found to be saturated, for an example, HD 19994, shown in the right panel of Figure \ref{images.fig}. To identify the saturated sources, we utilized the \texttt{photutils.profiles.radial\_profile} package to generate radial surface brightness profiles for each source. 
In the radial profile, if the brightness at a given radius is fainter than that at the outer radius, the source is flagged as a saturated source candidate, as shown in Figure \ref{profile.fig}. We then conducted a visual inspection to confirm the saturation cases.

GALEX provided averaged PSF images for the FUV and NUV bands, with the peak of each PSF scaled to unity. We did fifteenth-order polynomial fittings to the radial profiles of the PSFs. Figure \ref{psf_model.fig} shows the fitting results of the FUV and NUV bands, and Table \ref{psf.tab} lists the fitting coefficients. 

For each saturated star, we performed PSF photometry using the PSF provided by GALEX. First, we searched for the radius $r_0$ with the maximum brightness in the radial profile. Second, using the profile range from $r_0+3$ pixels to the outer edge, we matched the GALEX PSF to the observed radial profile. The adjusted PSF was then treated as the true radial profile of the saturated source.
Finally, we integrated the flux using the PSF and calculated the flux and magnitude using Equation \ref{cps_flux.eq} and \ref{cps_mag.eq}.

\subsubsection{Sources with Upper Limits}
\label{upperlimit.sec}

We presented upper limits for two types of sources:
(1) Sources fainter than the sensitivity limits of GALEX, which are approximately 19.9 AB mag in FUV and 20.8 AB mag in NUV for the AIS, and 22.7 AB mag in both bands for the MIS \citep{2017ApJS..230...24B}. (2) Sources with magnitude errors larger than 1. The photometric process for these sources is the same as those for the general sources. 

We compared the upper limit results with those calculated using the formula proposed by \cite{2013ApJ...766...60G}. They suggested that the upper limit magnitude, when the aperture magnitude is fainter than the upper limit, is a function of exposure time and sky background brightness, as follows,
\begin{equation}
    m_{\text{lim}} = -2.5 \log_{10} \left( 5 \sqrt{\frac{B_{\text{sky}} \cdot N_{\text{pix}}}{t_{\text{exp}}}} \right) + ZP + C_{\text{ap}},
\label{upperlimit.eq}
\end{equation}
where $m_{\text{lim}}$ represents the upper limit magnitude, \(B_{\text{sky}}\) is the sky background brightness with $B_{\text{sky}} = 3\times 10^{-3} {\rm counts \cdot s^{-1} \cdot pixel^{-1}}$, \(N_{\text{pix}}\) is the number of pixels, which is $16\pi$, \(t_{\text{exp}}\) is the exposure time, \(ZP\) is the zero point magnitude, and $C_{\text{ap}} = 0.23$ is the aperture correction.
Figure \ref{upperlimit.fig} shows the comparison for the FUV and NUV bands.
The two populations of upper limits estimated using Equation \ref{upperlimit.eq} are due to the different exposure times of the AIS and MIS surveys.
Our estimated upper limits are fainter compared to those obtained using the method by \cite{2013ApJ...766...60G}.

\begin{table*}
\centering
\caption{Photometric results of GALEX.}
\label{galex.tab}
\resizebox{\textwidth}{!}{
\begin{threeparttable}
\begin{tabular}{l*{14}{c}}
\hline\noalign{\smallskip}
Host name & $T_{\rm eff}$ & log$g$ & [Fe/H] & Ref.\tnote{1} & Distance & E(B-V) & Flag$_{\rm E(B-V)}$\tnote{2} & Stellar Type & $m_{\rm FUV}$ & Flag$_{\rm FUV}$\tnote{3} & $m_{\rm NUV}$ & Flag$_{\rm NUV}$\tnote{3} & log$R^{\prime}_{\rm FUV}$ & log$R^{\prime}_{\rm NUV}$ \\
  & (K) &  &  &  & (pc) &  &  &  & (mag) &  & (mag) &  &  &  \\
\hline\noalign{\smallskip}
11 Com & 4421.92 & 2.04 & -0.08 & 6 & 97.98 & 0.04 & 1 & Kg & $  19.45 \pm  0.04 $  & 0/0 & $  14.09 \pm  0.01 $  & 0/0 & -3.48 & -5.70 \\
14 Her & 5859.20 & 4.46 & -0.16 & 6 & 17.89 & 0.01 & 1 & Gd & $  20.71 \pm  0.08 $  & 0/0 & $  13.93 \pm  0.01 $  & 0/0 & -3.13 & -5.91 \\
17 Sco & 4347.11 & 1.91 & -0.06 & 0 & 136.57 & 0.00 & 0 & Kg & $  21.09 \pm  0.57 $  & 2/0 & $  15.21 \pm  0.01 $  & 0/0 & -4.28 & -6.71 \\
18 Del & 4812.70 & 2.70 & -0.13 & 6 & 74.94 & 0.00 & 0 & Kg & $  19.44 \pm  0.16 $  & 0/0 & $  13.40 \pm  0.01 $  & 0/0 & -3.32 & -5.80 \\
24 Boo & 5011.34 & 2.38 & -0.69 & 0 & 94.18 & 0.02 & 1 & Kg & $  19.93 \pm  0.18 $  & 2/0 & $  14.59 \pm  0.01 $  & 0/0 & -3.89 & -6.10 \\
24 Sex & 8294.70 & 3.58 & -0.09 & 6 & 73.14 & 0.00 & 0 & Ad & $  20.92 \pm  0.33 $  & 2/0 & $  13.96 \pm  0.01 $  & 0/0 & -4.00 & -6.86 \\
30 Ari B & 6851.35 & 4.31 & -0.33 & 6 & 44.37 & 0.00 & 0 & Fd & $  16.82 \pm  0.04 $  & 0/0 & $  13.29 \pm  0.01 $  & 0/0 & -2.75 & -4.24 \\
4 UMa & 4517.83 & 2.21 & -0.18 & 0 & 75.63 & 0.00 & 0 & Kg & $  19.76 \pm  0.19 $  & 0/0 & $  14.00 \pm  0.01 $  & 0/0 & -4.07 & -6.44 \\
42 Dra & 4446.59 & 2.05 & -0.41 & 0 & 90.25 & 0.00 & 0 & Kg & $  19.73 \pm  0.19 $  & 0/0 & $  14.11 \pm  0.01 $  & 0/0 & -4.10 & -6.42 \\
47 UMa & 6300.04 & 4.32 & -0.25 & 6 & 13.88 & 0.01 & 1 & Fd & $  17.10 \pm  0.05 $  & 0/0 & $  14.61 \pm  0.09 $  & 1/0 & -3.97 & -5.04 \\
51 Peg & 6208.55 & 4.34 & -0.22 & 6 & 15.51 & 0.00 & 0 & Fd & $  18.19 \pm  0.09 $  & 0/0 & $  13.28 \pm  0.01 $  & 0/0 & -3.33 & -5.37 \\
55 Cnc & 5743.90 & 4.47 & -0.12 & 6 & 12.58 & 0.00 & 0 & Gd & $  19.88 \pm  0.05 $  & 0/0 & $  13.46 \pm  0.01 $  & 0/0 & -3.26 & -5.89 \\
6 Lyn & 4950.66 & 3.12 & -0.13 & 0 & 54.81 & 0.00 & 0 & Kg & $  20.26 \pm  0.23 $  & 2/0 & $  13.48 \pm  0.01 $  & 0/0 & -3.17 & -5.95 \\
61 Vir & 5805.00 & 4.80 & 0.03 & 3 & 8.53 & 0.00 & 0 & Gd & $  17.74 \pm  0.06 $  & 0/1 & $  14.64 \pm  0.01 $  & 0/1 & -4.10 & -5.41 \\
70 Vir & 7438.68 & 4.18 & -0.13 & 6 & 18.09 & 0.00 & 0 & Fd & $  18.32 \pm  0.09 $  & 0/0 & $  15.31 \pm  0.13 $  & 1/0 & -4.74 & -6.01 \\
75 Cet & 4809.00 & 2.69 & 0.10 & 7 & 81.91 & 0.03 & 1 & Kg & $  20.17 \pm  0.06 $  & 0/0 & $  13.46 \pm  0.01 $  & 0/0 & -3.30 & -6.06 \\
8 UMi & 5148.47 & 2.71 & -0.46 & 6 & 162.68 & 0.00 & 0 & Kg & $  21.71 \pm  0.52 $  & 2/0 & $  14.72 \pm  0.01 $  & 0/0 & -3.43 & -6.29 \\
81 Cet & 4770.80 & 2.46 & -0.12 & 6 & 104.69 & 0.03 & 1 & Kg & $  20.06 \pm  0.20 $  & 2/0 & $  13.86 \pm  0.01 $  & 0/0 & -3.30 & -5.85 \\
AB Aur & 9770.00 & 3.77 & -999 & 7 & 155.05 & 0.00 & 0 & Ad & $  12.26 \pm  0.01 $  & 0/0 & $  14.65 \pm  0.09 $  & 1/0 & -3.47 & -2.59 \\
AB Pic & 4749.80 & 4.50 & -0.09 & 0 & 50.13 & 0.05 & 1 & Kd & $  19.37 \pm  0.15 $  & 0/0 & $  15.54 \pm  0.01 $  & 0/0 & -2.44 & -4.06 \\
\hline\noalign{\smallskip}
\end{tabular}
\begin{tablenotes}
        \normalsize 
        \item[1] References of atmospheric parameters, with 0: APOGEE DR17; 1: LAMOST DR 10 LRS; 2: LAMOST DR 10 MRS; 3: RAVE DR6; 4: GALAH DR3; 5: StarHorse \citep{2023A&A...673A.155Q}; 6: StarHorse \citep{2019A&A...628A..94A}; 7: \cite{ps}.
        \item[2] Flags for extinction values, with 0: PS1 3D dust map; 1: SFD extinction; 2: fixed value of 0.1 for stars with distances less than 200 pc but with high SFD extinction.
        \item[3] Flags for FUV and NUV photometry. The numbers before the `/' represent photometry types, with 0: general photometry; 1: saturated stars; 2: (undetected) stars with upper limits. The numbers after the `/' represent the background measurement methods, with 0: estimation from GALEX background maps; 1: interpolation with GALEX images. 
      \end{tablenotes}
      This table is available in its entirety in machine-readable and Virtual Observatory (VO) forms in the online journal. A portion is shown here for guidance regarding its form and content.
    \end{threeparttable}
}
\end{table*}

\begin{table*}
\centering
\caption{Photometric results from a single observation of GALEX.}
\label{galex_multi.tab}
\resizebox{\textwidth}{!}{
\begin{threeparttable}
\begin{tabular}{l*{6}{c}}
\hline\noalign{\smallskip}
Host name & $t_{\rm obs,FUV}$\tnote{1} & $m_{\rm FUV}$ & Flag$_{\rm FUV}$\tnote{2} & $t_{\rm obs,NUV}$\tnote{1} & $m_{\rm NUV}$ & Flag$_{\rm NUV}$\tnote{2} \\
  & (s) & (mag) &  & (s) & (mag) &   \\
\hline\noalign{\smallskip}
11 Com & 1174631415.495 & $  19.59 \pm  0.17 $  & 0/0 & 1174631415.495 & $  12.97 \pm  0.01 $  & 0/0 \\
11 Com & 1178991380.495 & $  19.51 \pm  0.04 $  & 0/0 & 1178991380.495 & $  15.48 \pm  0.01 $  & 0/0 \\
14 Her & 1242849139.995 & $  20.71 \pm  0.08 $  & 0/0 & 1242849139.995 & $  13.94 \pm  0.01 $  & 0/0 \\
17 Sco & 1181648539.995 & $  21.09 \pm  0.57 $  & 2/0 & 1181648539.995 & $  15.19 \pm  0.01 $  & 0/0 \\
17 Sco & 1184050110.495 & $  21.20 \pm  0.71 $  & 2/0 & 1184050110.495 & $  15.15 \pm  0.01 $  & 0/0 \\
18 Del & 1158023254.495 & $  19.44 \pm  0.16 $  & 0/0 & 1158023254.495 & $  13.41 \pm  0.01 $  & 0/0 \\
18 Del & 1158129775.995 & $  19.09 \pm  0.14 $  & 0/0 & 1158129775.995 & $  13.55 \pm  0.01 $  & 0/0 \\
24 Boo & 1176832357.495 & $  19.91 \pm  0.19 $  & 2/0 & 1176832357.495 & $  14.91 \pm  0.01 $  & 0/0 \\
24 Sex & 1201431622.995 & $  20.92 \pm  0.33 $  & 2/0 & 1201431622.995 & $  13.98 \pm  0.01 $  & 0/0 \\
24 Sex & 1143196404.495 & $  21.07 \pm  0.36 $  & 2/0 & 1143196404.495 & $  13.96 \pm  0.01 $  & 0/0 \\
2MASS J01225093-2439505 & 1161011448.495 & $  21.19 \pm  0.36 $  & 2/0 & 1161011448.495 & $  20.70 \pm  0.17 $  & 2/0 \\
2MASS J01225093-2439505 & 1164649824.995 & $  21.49 \pm  0.53 $  & 2/0 & 1164649824.995 & $  20.48 \pm  0.18 $  & 2/0 \\
2MASS J02192210-3925225 & 1132517083.495 & $  22.05 \pm  0.49 $  & 2/1 & 1132517083.495 & $  21.32 \pm  0.26 $  & 2/1 \\
2MASS J22362452+4751425 & 1215950236.495 & $  24.62 \pm  10.37 $  & 2/0 & 1215950236.495 & $  20.57 \pm  0.18 $  & 2/0 \\
2MASS J22362452+4751425 & 1215914727.495 & $  21.45 \pm  0.66 $  & 2/0 & 1215914727.495 & $  21.96 \pm  0.55 $  & 2/0 \\
30 Ari B & 1193148388.995 & $  16.81 \pm  0.05 $  & 0/0 & 1193148388.995 & $  13.81 \pm  0.01 $  & 0/0 \\
30 Ari B & 1193757753.495 & $  16.85 \pm  0.05 $  & 0/0 & 1193757753.495 & $  13.72 \pm  0.01 $  & 0/0 \\
30 Ari B & 1227001096.995 & $  16.81 \pm  0.05 $  & 0/0 & 1227001096.995 & $  13.83 \pm  0.01 $  & 0/0 \\
30 Ari B & 1227101693.495 & $  16.82 \pm  0.04 $  & 0/0 & 1227101693.495 & $  13.30 \pm  0.01 $  & 0/0 \\
4 UMa & 1167104845.495 & $  19.76 \pm  0.19 $  & 0/0 & 1167104845.495 & $  14.01 \pm  0.01 $  & 0/0 \\
\hline\noalign{\smallskip}
\end{tabular}
\begin{tablenotes}
        \normalsize 
        \item[1] Observational times for the FUV and NUV bands, calculated as the mean values of the start and end times. GALEX TIME = UNIX TIME - 315964800 \citep{2016ApJ...833..292M}.
        \item[2] Flags for FUV and NUV photometry. The numbers before the `/' represent photometry types, with 0: general photometry; 1: saturated stars; 2: (undetected) stars with upper limits. The numbers after the `/' represent the background measurement methods, with 0: estimation from GALEX background maps; 1: interpolation with GALEX images. 
      \end{tablenotes}
      This table is available in its entirety in machine-readable and Virtual Observatory (VO) forms in the online journal. A portion is shown here for guidance regarding its form and content.
    \end{threeparttable}
}
\end{table*}

\section{SWIFT photometry}
\label{swift.sec}

The Swift UVOT is one of three instruments onboard the Swift satellite, launched by NASA in 2004. UVOT is designed to provide simultaneous UV and optical imaging and photometry, operating over a wavelength range of 170–600 nm. It has high spatial resolution (about 2.5\arcsec/pixel for the UV filters) and a wide field of view (17$\times$17 square arcminutes) \citep{2008MNRAS.383..627P}. UVOT includes six filters, i.e., three in the UV range and three in the optical range. The UV filters include UVW2 (1800-2600 \AA), UVM2 (2000-2800 \AA) and UVW1 (2400-3400 \AA) \citep{2008MNRAS.383..627P}.

\subsection{Data preparation}
\label{swift_img.sec}

To supplement GALEX photometry, we searched for sources observed by SWIFT. The search was performed using the Swift UVOT data archive\footnote{\url{https://archive.stsci.edu/swiftuvot/search.php}} with a search radius of 0.5\arcmin. This provided a list of available observations in the UVOT bands. We focused only on the UVW2, UVM2, and UVW1 bands for our analysis. In total, using \texttt{wget}, we downloaded 74 images in the UVM2 band, 95 images in the UVW2 band, and 34 images in the UVW1 band for 27 host stars.

\subsection{Photometry steps}
\label{swift_pho.sec}

We performed photometry using individual exposures.
First, we checked each image and excluded those due to the presence of large ghost images that typically overlap with these saturated sources. As a result, we did not conduct PSF photometry for SWIFT observations.
Second, we corrected the coordinates of each target to match their positions at the time of observations, using the proper motion data from the Gaia DR3.
Third, we calculated the corrected exposure times for each observation following \cite{2008MNRAS.383..627P}.
Forth, we determined the background values using two methods.
We created an annulus around the star, with an inner radius of 55 pixels and an outer radius of 70 pixels, to calculate the counts per pixel in this annulus. 
When this value is larger than 10, we created the background map following the method described in Section \ref{galex_pho.sec}. 
When this value is smaller than 10, we estimated the background using this annulus as suggested by \citep{2008MNRAS.383..627P}.
Fifth, we did the aperture photometry using a radius of 10 pixels (5\arcsec).
Finally, we converted the net count rate (i.e., background-subtracted) to fluxes and magnitudes following Equation \ref{cps_flux.eq} and \ref{cps_mag.eq}. For the UM2, UW2, and UW1 bands, the constants $C1$ are $8.489\times10^{-16}$, $6.225\times10^{-16}$, and $4.623\times10^{-16}$, respectively; the constants $C$2 are 18.54, 19.11, and 18.95, respectively\footnote{\url{https://heasarc.gsfc.nasa.gov/docs/heasarc/caldb/swift/docs/uvot/uvot_caldb_AB_10wa.pdf}}. 
The photometry results of SWIFT are summarized in Table \ref{swift.tab}.

\begin{table*}
\centering
\caption{Photometric results of SWIFT.}
\label{swift.tab}
\resizebox{\textwidth}{!}{
\begin{threeparttable}
\begin{tabular}{l*{10}{c}}
\hline\noalign{\smallskip}
Host name & $T_{\rm eff}$ & log$g$ & [Fe/H] & Ref.\tnote{1} & Distance & E(B-V) & Flag$_{\rm E(B-V)}$\tnote{2} & $m_{\rm UW2}$ & $m_{\rm UM2}$ & $m_{\rm UW1}$ \\
& (K) &  &  &  & (pc) & & & (mag) & (mag) & (mag) \\
\hline\noalign{\smallskip}
55 Cnc & 5743.90 & 4.47 & -0.12 & 6 & 12.58 & 0.00 & 0 &  -999    & $ 14.03 \pm  0.01 $ &  -999    \\
AU Mic & 3655.12 & 4.46 & 0.14 & 6 & 9.71 & 0.06 & 1 &  -999    & $ 16.15 \pm  0.01 $ &  -999    \\
GJ 15 A & 3700.28 & 4.70 & -0.47 & 0 & 3.56 & 0.00 & 0 &  -999    &  -999    &  -999    \\
GJ 229 & 3582.71 & 4.69 & 0.17 & 6 & 5.76 & 0.00 & 0 & $ 15.55 \pm  0.01 $ &  -999    &  -999    \\
GJ 581 & 3222.83 & 4.88 & 0.04 & 6 & 6.30 & 0.00 & 0 &  -999    &  -999    & $ 16.85 \pm  0.02 $ \\
GJ 674 & 3279.42 & 4.85 & 0.09 & 6 & 4.55 & 0.10 & 2 &  -999    &  -999    & $ 15.73 \pm  0.01 $ \\
GJ 832 & 3449.99 & 4.77 & -0.04 & 6 & 4.97 & 0.03 & 1 &  -999    & $ 19.38 \pm  0.04 $ &  -999    \\
HAT-P-67 & 6940.78 & 3.83 & -0.14 & 6 & 370.06 & 0.01 & 0 &  -999    & $ 14.81 \pm  0.01 $ &  -999    \\
HD 141004 & 5885.38 & 4.17 & 0.03 & 7 & 11.90 & 0.03 & 1 &  -999    & $ 13.50 \pm  0.01 $ &  -999    \\
HD 179949 & 6131.85 & 4.30 & -0.05 & 6 & 27.51 & 0.00 & 0 & $ 14.21 \pm  0.01 $ &  -999    &  -999    \\
HD 189733 & 5034.09 & 4.58 & 0.01 & 6 & 19.76 & 0.00 & 0 & $ 14.59 \pm  0.01 $ &  -999    &  -999    \\
HD 192310 & 5352.47 & 4.55 & -0.05 & 6 & 8.81 & 0.00 & 0 &  -999    & $ 14.03 \pm  0.01 $ &  -999    \\
HD 69830 & 5385.00 & 4.49 & -0.05 & 7 & 12.58 & 0.00 & 0 &  -999    & $ 13.72 \pm  0.01 $ &  -999    \\
HD 7924 & 5131.00 & 4.46 & -0.21 & 7 & 17.00 & 0.00 & 0 &  -999    & $ 14.76 \pm  0.01 $ &  -999    \\
HU Aqr & 5952.00 & -999 & -999 & 7 & 189.60 & 0.00 & 0 &  -999    & $ 15.73 \pm  0.01 $ &  -999    \\
KELT-11 & 5370.71 & 3.59 & -0.36 & 6 & 99.63 & 0.00 & 0 &  -999    & $ 15.50 \pm  0.01 $ &  -999    \\
Kepler-293 & 5581.08 & 4.53 & -0.12 & 6 & 959.19 & 0.04 & 0 &  -999    &  -999    &  -999    \\
LHS 1140 & 3096.00 & 5.04 & -0.15 & 7 & 14.96 & 0.02 & 1 &  -999    &  -999    &  -999    \\
MXB 1658-298 & -999 & -999 & -999 & 7 & 6024.51 & 0.42 & 0 &  -999    &  -999    &  -999    \\
Proxima Cen & 2900.00 & 5.16 & -999 & 7 & 1.30 & 0.10 & 2 &  -999    &  -999    & $ 18.36 \pm  0.18 $ \\
Teegarden's Star & 3034.00 & 5.19 & -0.11 & 7 & 3.83 & 0.00 & 0 &  -999    &  -999    &  -999    \\
WASP-121 & 6757.49 & 4.20 & -0.13 & 6 & 262.40 & 0.18 & 1 &  -999    &  -999    & $ 14.38 \pm  0.01 $ \\
WASP-18 & 6611.00 & 4.47 & 0.29 & 3 & 122.61 & 0.01 & 1 &  -999    &  -999    & $ 14.00 \pm  0.01 $ \\
WASP-76 & 6250.00 & 4.13 & 0.23 & 7 & 188.04 & 0.03 & 0 &  -999    & $ 15.03 \pm  0.01 $ &  -999    \\
WASP-79 & 6471.00 & 3.94 & -0.29 & 3 & 244.70 & 0.03 & 1 &  -999    &  -999    & $ 14.12 \pm  0.01 $ \\
WASP-82 & 6441.66 & 3.99 & 0.15 & 1 & 273.35 & 0.07 & 0 &  -999    & $ 15.06 \pm  0.01 $ &  -999    \\
ups And & 6090.64 & 4.16 & 0.11 & 0 & 13.47 & 0.00 & 0 & $ 14.28 \pm  0.01 $ &  -999    &  -999    \\
\hline\noalign{\smallskip}
\end{tabular}
\begin{tablenotes}
        \normalsize 
        \item[1] References of atmospheric parameters, with 0: APOGEE DR17; 1: LAMOST DR 10 LRS; 2: LAMOST DR 10 MRS; 3: RAVE DR6; 4: GALAH DR3; 5: StarHorse \citep{2023A&A...673A.155Q}; 6: StarHorse \citep{2019A&A...628A..94A}; 7: \cite{ps}.
        \item[2] Flags for extinction values, with 0: PS1 3D dust map; 1: SFD extinction; 2: fixed value of 0.1 for stars with distances less than 200 pc but with high SFD extinction.
      \end{tablenotes}
      This table is available in its entirety in machine-readable and Virtual Observatory (VO) forms in the online journal.
    \end{threeparttable}
}
\end{table*}

\section{Habitability}
\label{hz.sec}

Several confirmed exoplanets, such as Kepler-186f and Proxima Centauri b, have been reported within their star's HZ \citep{2014Sci...344..277Q, 2016Natur.536..437A}. For example, Kepler-186 f orbits a cool M-dwarf star and receives a similar amount of stellar flux as Mars, placing it at the outer edge of the HZ \citep{2014Sci...344..277Q}. Proxima Centauri b, the closest known exoplanet, resides within the HZ of its parent star but suffers from tidal locking, stellar flares, and strong UV and X-ray emission \citep{2016Natur.536..437A}.
In this section, we will make a comprehensive study of the habitability of planets orbiting stars with GALEX NUV photometry.

\subsection{Physical Parameters}
\label{pars.sec}

In order to study habitability and stellar activity, we first need to obtain the parameters of both stars and their planets.

\subsubsection{Parameters of host stars}
\label{stars_pars.sec}

The catalog of Planetary Systems Composite Data (PSCD) from \cite{ps} provides abundant information of each planet and its host star, including stellar physical parameters, atmospheric parameters, distances, etc. However, the atmospheric parameters for each star in this catalog were collected from various references, and some may be unreliable. To obtain accurate atmospheric parameters for host stars, we prioritized using data from spectral surveys, supplementing them with the \textit{StarHorse} catalog \citep{2018MNRAS.476.2556Q} and the PSCD catalog. 
\textit{StarHorse} is a Bayesian isochrone-fitting tool for determining stellar atmospheric parameters, distances and extinctions by combining spectroscopic, photometric, and astrometric data.

We cross-matched our sample (with GALEX NUV photometry) with several major spectroscopic surveys: Apache Point Observatory Galactic Evolution Experiment Data Release 17\footnote{\url{https://www.sdss4.org/dr17/irspec/spectro_data/}} (APOGEE DR17), Large Sky Area Multi-Object Fiber Spectroscopic Telescope Data Release 10\footnote{\url{http://www.lamost.org/dr10/v1.0/}} (LAMOST DR10), RAdial Velocity Experiment Data Release 6\footnote{\url{https://www.rave-survey.org/}} (RAVE DR6), and Galactic Archaeology with HERMES Data Release 3\footnote{\url{https://www.galah-survey.org/dr3/overview/}} (GALAH DR3). The match radius is 3\arcsec. For each survey, we used a S/N weighted average method to calculate the most reliable atmospheric parameters for each host star with multiple observations, following 
\begin{equation} \label{eqweight}
\overline{P} = \frac{\sum_k w_k \cdot P_{k}}{\sum_k w_k}
\end{equation}
and
\begin{equation}
\sigma_w(\overline{P}) = \sqrt{\frac{N}{N-1}\frac{\sum_k w_k \cdot (P_{k} - \overline{P})^2}{\sum_k w_k}}.
\end{equation}
The index $k$ is the epoch of the measurements of parameter $P$, and the weight $w_k$ is estimated as the square of the S/N for each spectrum \citep{2024ApJ...966...69L}. We then selected atmospheric parameters from different surveys according to the sequence of APOGEE, LAMOST, RAVE, and GALAH.

Additionally, we supplemented the atmospheric parameters (i.e., $T_{\text{eff}}$, $\log g$, and [Fe/H]) from \textit{StarHorse}. We cross-matched our sample with the catalogs from \cite{2023A&A...673A.155Q} and \cite{2019A&A...628A..94A} within a 5\arcsec\ radius.

We also compared the atmospheric parameters from different catalogs with those in the catalog from \cite{ps}, as shown in Figure \ref{pars.fig}. 
The comparison reveals that the effective temperature and surface gravity are in good agreement, while the metallicity distribution is diffuse.

Gaia eDR3 provided distance measurements for approximately 1.47 billion objects \citep{2021AJ....161..147B}. We cross-matched the host star sample with the Gaia eDR3 distance catalog using a match radius of 3\arcsec, and supplemented it with the PSCD catalog.

The reddening $E(B-V)$ was derived from the Pan-STARRS DR1 (PS1) 3D dust map with $E(B-V) =0.884 \times {\rm Bayestar19}$, the latter being a measure of extinction given by \cite{2019ApJ...887...93G}.
For sources without extinction estimation from the PS1 dust map, we used the SFD dust map with $E(B-V) =0.884 \times E(B-V)_{\rm SFD}$ as a supplement. SFD dust map provides two-dimensional measurements of dust extinction in the Milky Way based on far-infrared emissions observed by the IRAS and COBE satellites \citep{1998ApJ...500..525S}. 
For nearby sources, the SFD extinction values may be less accurate. Thus, we used a reddening value of 0.1 for sources within 200 pc.

\subsubsection{Parameters of exoplanets}
\label{planets_pars.sec}

The PSCD catalog provides parameters of each planet, including planet radius, planet mass, orbital semi-major axis, and orbital period, etc. These parameters can help us to make simple classification of planets and study their habitability.

We also simply classified the exoplanets using their masses and radii. Gas giants were defined with masses greater than $20\ M_\oplus$ and radii exceeding $4\ R_\oplus$. Neptune-like planets were classified as those with masses between 2 and 20 $M_\oplus$, and radii between 2.5 and 4 $R_\oplus$. Super-Earths were identified with masses between 1 and 10 $M_\oplus$ and radii between 1.5 and 2.5 $R_\oplus$.  Terrestrials were defined as those with masses ranging from 0.1 to 10 $M_\oplus$, and radii ranging from 0.5 to 1.5 $R_\oplus$.

\subsection{Habitable Zone Calculation}
\label{chz.sec}

We derived the CHZs of the host stars using the method described by \citet{2014ApJ...787L..29K}. The inner edge of CHZ was calculated based on the ``runaway greenhouse" limit (i.e., the greenhouse effect caused by water), and the outer edge was determined using the ``maximum greenhouse" limit. 
The CHZ was calculated following  
$d=(\frac{L/L_{\odot}}{S_{\rm eff}})^{0.5}$ AU \citep{2014ApJ...787L..29K},
where $S_{\rm eff}$ is the effective solar flux incident on the planet. For these calculations, we assumed a planet mass equal to that of Earth. We also calculated the CHZ for planets with masses equal to 0.1 and 5 times that of Earth. The results showed only minor differences in the inner edge of the CHZ \citep{2014ApJ...787L..29K}. Therefore, we used the parameters of an Earth-mass planet in all CHZ calculations.

Additionally, we calculated the boundaries of UHZ defined by \cite{2023MNRAS.522.1411S} for sources with GALEX NUV photometry. The outer boundary of the UHZ was established where $f_{\rm UV} \geq$ 45 erg cm$^{-2}$ $s^{-1}$, which is the minimum NUV flux required for abiogenesis \citep{2018SciA....4.3302R}, and the inner boundary was set with $f_{\rm UV} \leq$ 1.04$\times10^4$ erg cm$^{-2}$ $s^{-1}$, which is twice the intensity of UV radiation received by the Archean Earth approximately 3.8 billion years ago. 
We also considered the variations in flux reaching the planetary surface under different atmospheric absorption conditions. The flux can be reduced to 10\%, 50\%, and 100\% of that at the top of the atmosphere, corresponding to a transparency of the atmosphere to UV radiation $f$ of 0.1, 0.5, and 1.0, respectively.

\subsection{Habitability Distribution}
\label{hz_discuss.sec}

\begin{figure*}[htp]
    \centering
    \subfigure[]{
    \includegraphics[width=0.48\textwidth]{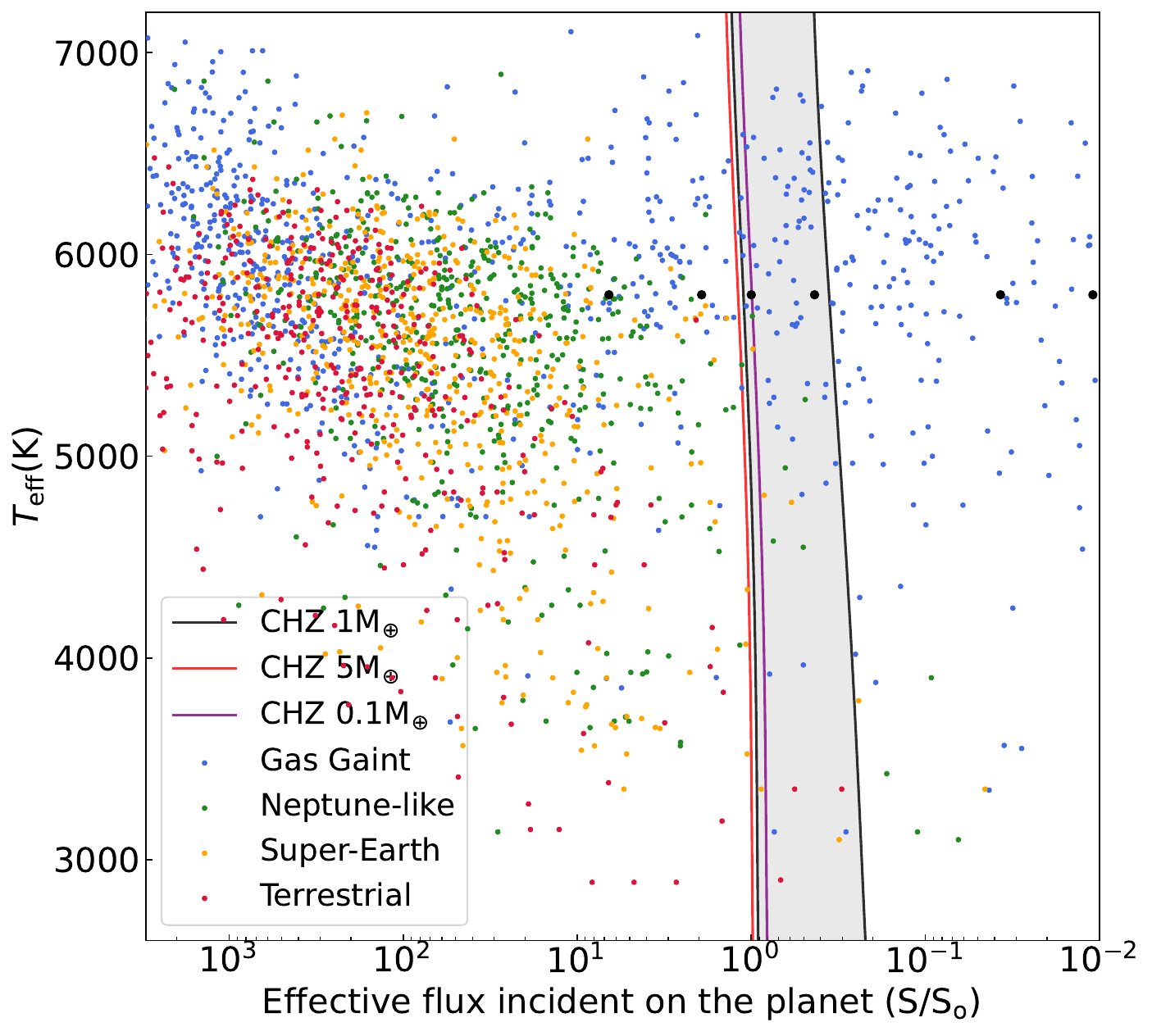}
    \label{chz.fig}}
    \subfigure[]{
    \includegraphics[width=0.48\textwidth]{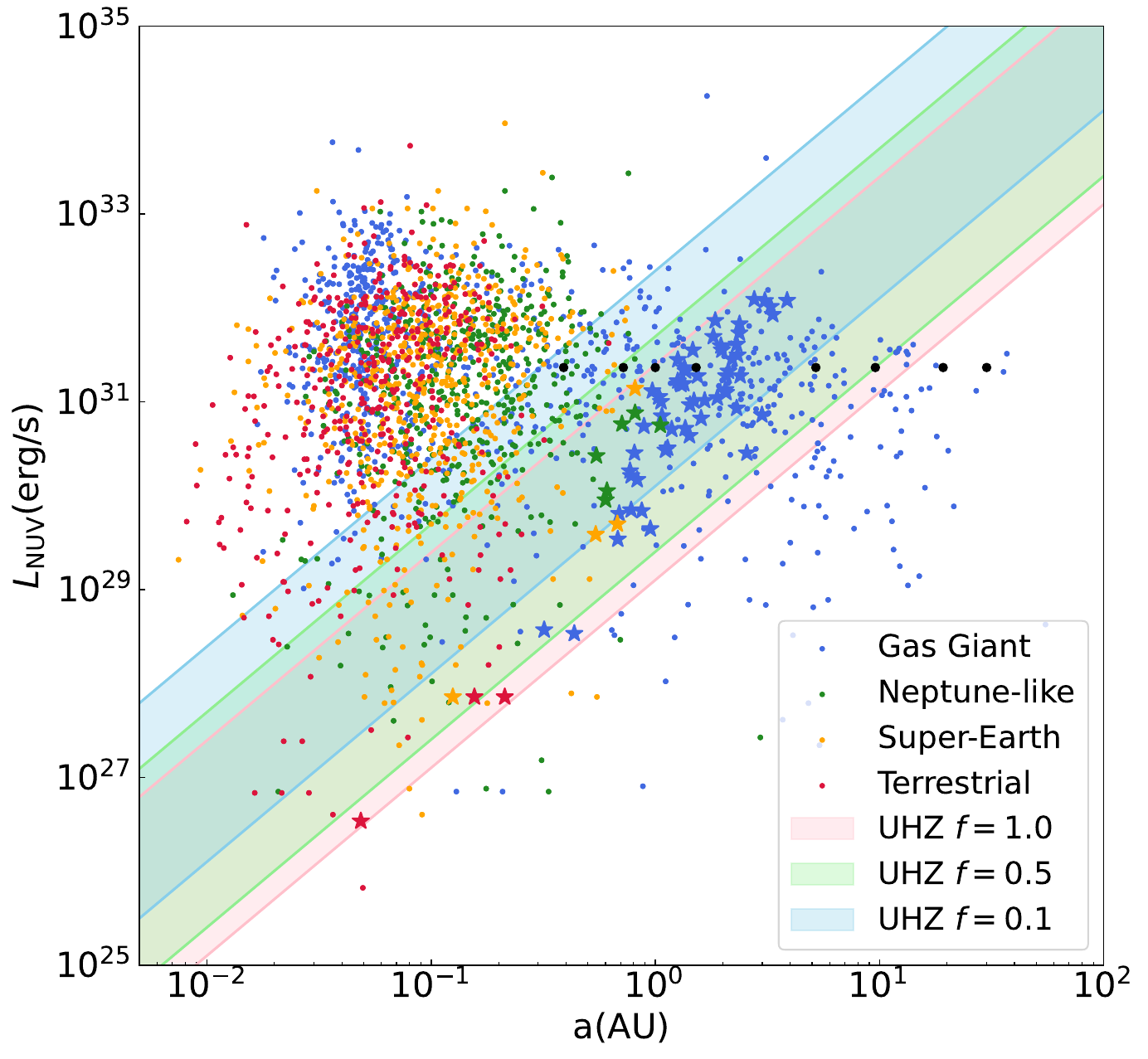}
    \label{uhz.fig}}
    \caption{Left panel: The relation between the effective temperatures of the host stars and the effective fluxes incident on their planets. The blue, green, orange, and red dots represent gas giants, Neptunian-like planets, super-Earths, and terrestrial planets, respectively. The gray shaded area is the CHZ for each star, and the black, red, and purple lines represent the inner and outer boundaries of the CHZ for Earth mass, 5 Earth masses, and 0.1 Earth masses, respectively. Note that the three outer boundary lines overlap. The black dots represent the planets in our Solar System. Right panel: the relation between the NUV luminosities of the host stars and the distances between the stars and their planets. The pink, green, and blue shaded areas represent the UHZ for different NUV transmission rates of planetary atmospheres. The blue, green, orange, and red asterisks represent gas giants, Neptunian-like planets, super-Earths, and terrestrial planets that are located within both the CHZ and UHZ.}  
    \label{hz.fig}
\end{figure*}

Figure \ref{hz.fig} shows the CHZ and UHZ of our sample. On the one hand, the majority of discovered exoplanets are not located within the CHZ but are situated very close to their host stars. Only 86 planets are within the CHZ, including 63 Gas Giants, 9 Neptunian-like planets, 6 Super-Earths, and 3 Terrestrial planets. The three Terrestrial planets are Proxima Cen b, GJ 667 C e, and GJ 667 C f. 
We also calculated the flux received by the Solar System planets for comparison. Earth and Mars are within the CHZ, with Earth located near the inner edge of the CHZ \citep{article}. On the other hand, the majority of the planets are also not located within the UHZ. 
The overlapping region between CHZ and UHZ is considered as more favorable for habitability. We highlighted the planets located within both the CHZ and UHZ. The habitability of each planet is listed in Table \ref{hz.tab}.

\begin{table}
\resizebox{\columnwidth}{!}{%
\begin{threeparttable}
\centering
\caption{The habitability assessment of each planet.}
\label{hz.tab}
\begin{tabular}{l*{4}{c}}
\hline\noalign{\smallskip}
Planet & Radius & Mass & Type & HZ flag$^1$ \\
 & ($R_{\oplus}$) & ($M_{\oplus}$) &  &  \\
\hline\noalign{\smallskip}
14 Her b & 12.60 & 2559.47 & Gas Giant & 2 \\
16 Cyg B b & 13.50 & 565.74 & Gas Giant & 3 \\
24 Sex b & 13.40 & 632.46 & Gas Giant & 2 \\
24 Sex c & 13.90 & 273.32 & Gas Giant & 2 \\
30 Ari B b & 12.30 & 4392.41 & Gas Giant & 0 \\
51 Peg b & 14.30 & 146.20 & Gas Giant & 0 \\
55 Cnc b & 13.90 & 263.98 & Gas Giant & 0 \\
55 Cnc c & 8.51 & 54.47 & Gas Giant & 0 \\
55 Cnc d & 13.00 & 1232.49 & Gas Giant & 0 \\
55 Cnc e & 1.88 & 7.99 & Super-Earth & 0 \\
55 Cnc f & 7.59 & 44.81 & Gas Giant & 3 \\
61 Vir b & 2.11 & 5.10 & Super-Earth & 0 \\
61 Vir c & 4.46 & 18.20 & Neptune-like & 2 \\
61 Vir d & 5.11 & 22.90 & Neptune-like & 2 \\
AB Pic b & 12.30 & 4290.50 & Gas Giant & 0 \\
AU Mic b & 3.96 & 20.12 & Neptune-like & 0 \\
AU Mic c & 3.24 & 9.60 & Neptune-like & 2 \\
BD+14 4559 b & 13.80 & 330.54 & Gas Giant & 3 \\
BD+20 594 b & 2.58 & 22.25 & Neptune-like & 0 \\
BD+45 564 b & 13.60 & 432.25 & Gas Giant & 3 \\
\hline\noalign{\smallskip}
\end{tabular}
\begin{tablenotes}
        \normalsize 
        \item[1] Flags for habitability, with 0: neither in CHZ nor in UHZ; 1: in CHZ but not in UHZ; 2: in UHZ but not in CHZ; 3: in both CHZ and UHZ.
      \end{tablenotes}
      This table is available in its entirety in machine-readable and Virtual Observatory (VO) forms in the online journal.  A portion is shown here for guidance regarding its form and content.
    \end{threeparttable}
}
\end{table}

\begin{figure*}[htp]
    \centering
    \subfigure[]{
    \includegraphics[width=0.48\textwidth]{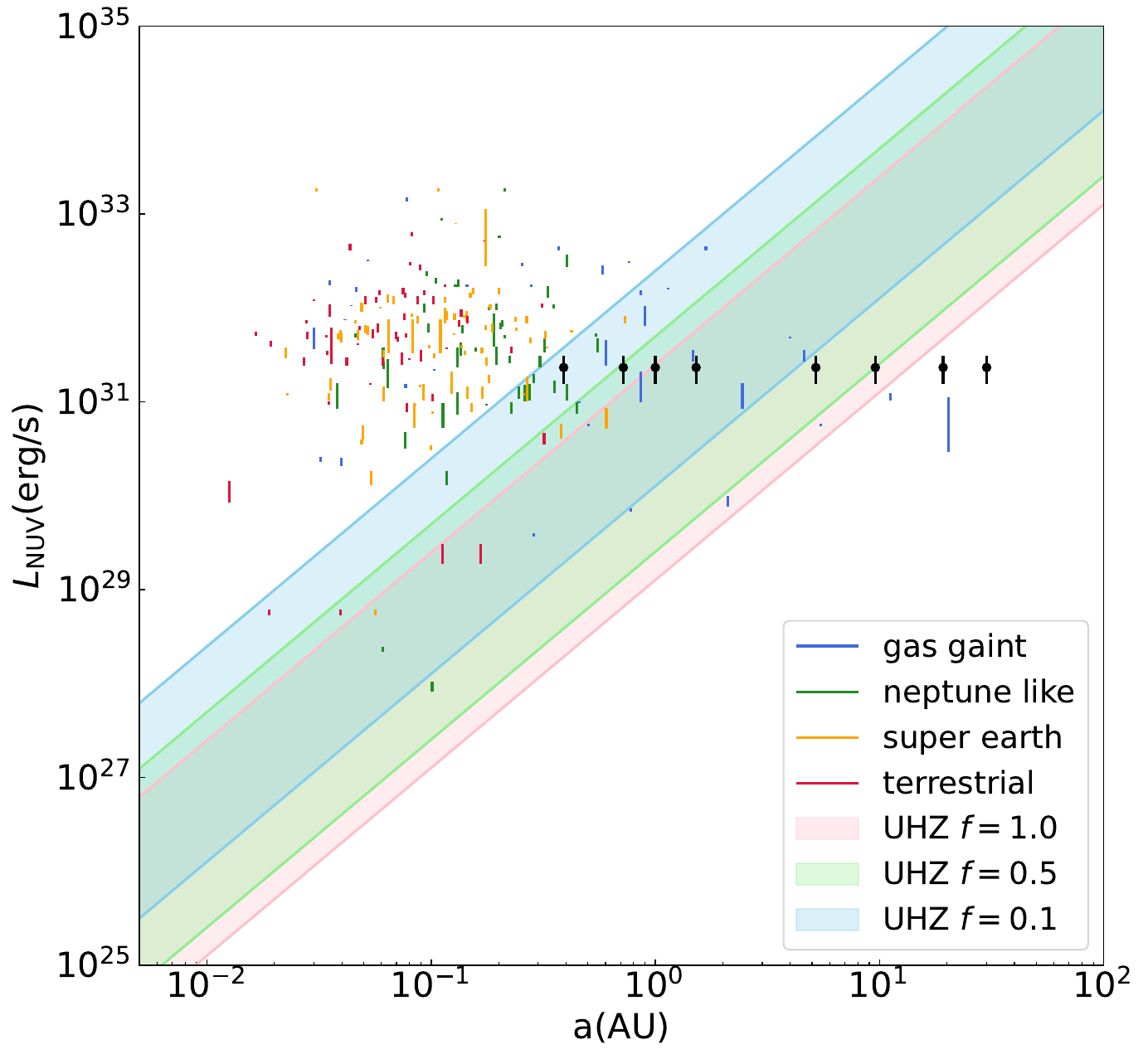}
    \label{uhz_variation.fig}}
    \subfigure[]{
    \includegraphics[width=0.48\textwidth]{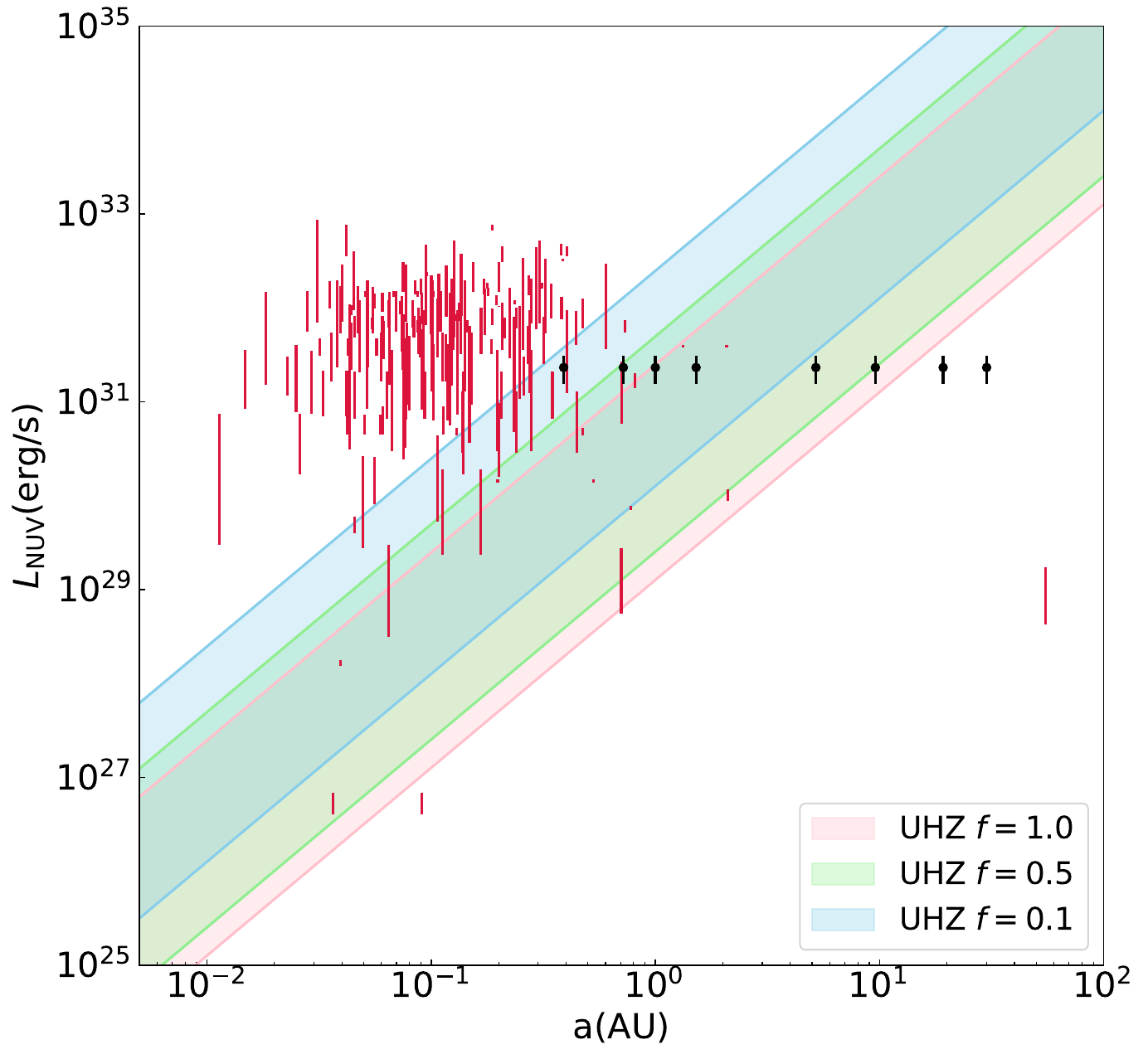}
    \label{uhz_flare.fig}}
    \caption{Left panel: The effect of stellar NUV luminosity variation on the UHZ. The symbols are the same as in Figure \ref{uhz.fig}. The line lengths indicate the amplitude of luminosity variation during multiple GALEX NUV observations. Right panel: The effect of stellar NUV luminosity variation caused by flares on the UHZ. }
    \label{variation.fig}
\end{figure*}

We also examined the impact of stellar UV emission variability on the UHZ for stars with multiple observations, as shown in Figure \ref{uhz_variation.fig}.  We selected stars with more than 10 NUV observations and calculated variability as the difference between the 95th and the 5th percentile of their brightness. 
The longest time span for the multiple observations is about 3000 days.
We found that the UV variability of these host stars has a minor impact on the UV habitability of their planets. 

Furthermore, we studied the impact of flares on the UHZ. For each observation, we used \textit{gPhoton} to bin the data every 20 seconds and performed aperture photometry on each bin. We identified a flare if there were at least three consecutive data points exceeding 3$\sigma$, where $\sigma$ is the standard deviation of the light curve. After visual inspection, we found a total of 157 flares in the NUV band. For these flares, we calculated their peak luminosities and used these values as the NUV luminosities to investigate the variability of the UHZ (Figure \ref{uhz_flare.fig}). Although flares can significantly enhance stellar UV radiation, potentially pushing some sources from habitable to uninhabitable, we think the overall impact on the UHZ is small due to the short duration of flares.  We also identified flare events in the FUV band and calculated their peak luminosities and energies. The calculation for the flares in both FUV and NUV bands are detailed in the appendix \ref{flare.sec}, and the results for the flares are presented in the Table \ref{flare.tab}.


We also estimated solar UV variability using data from Solar Stellar Irradiance Comparison Experiment (SOLSTICE), a space-based instrument designed to measure solar UV radiation \citep{2017JSWSC...7A...6L}. 
The wavelength range of the SOLSTICE Middle UV (MUV) band is about 1800--3100 \AA\footnote{\url{https://lasp.colorado.edu/sorce/data/}}, similar to the GALEX NUV wavelength range (1771--2831 \AA). Therefore, we used the SOLSTICE MUV data from May 2003 to October 2015, covering the descending phase of Cycle 23 and the rising phase of Cycle 24 \citep{2017JSWSC...7A...6L}, to trace the solar variation in the NUV band. 
Figure \ref{uhz_variation.fig} shows the NUV luminosity variation received by solar planets over about 12 years. 
The NUV variation experienced by these planets is very small, similar to what is observed for other exoplanets.

To investigate the fate of the planets, we compared their distances to their host stars with the maximum radii of the host stars when they evolve into the red giant stages. We used the MIST EEP Tracks models \footnote{\url{https://waps.cfa.harvard.edu/MIST/model_grids.html}}, with [Fe/H] ranges from $-1.5$ to 0.5 in 0.25 dex steps at $v/v_{\rm crit}=0$, to analyze the evolutionary tracks of stars. 
Using the closest metallicity and stellar mass (derived from the PSCD catalog), we selected the best-fit stellar track and obtained the maximum stellar radii during the red giant phase (i.e., phase$=$2 in MIST models).
This allowed us to predict which planets would be engulfed by their host stars due to stellar evolution. Here we did not take into account planetary migration effects.

\begin{figure*}[htp]
    \centering
    \subfigure[]{
    \includegraphics[width=0.48\textwidth]{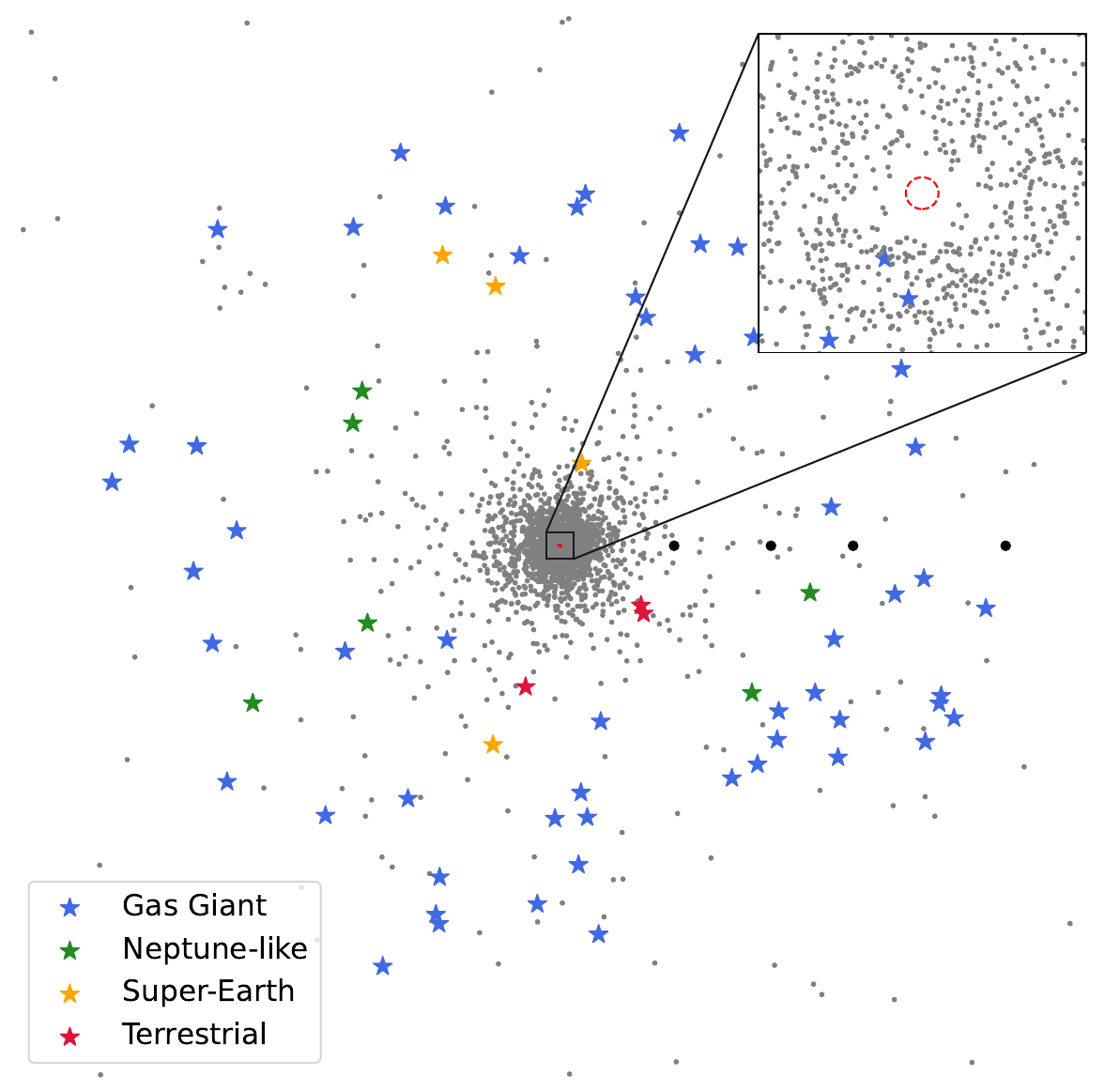}
    \label{ms.fig}}
    \subfigure[]{
    \includegraphics[width=0.48\textwidth]{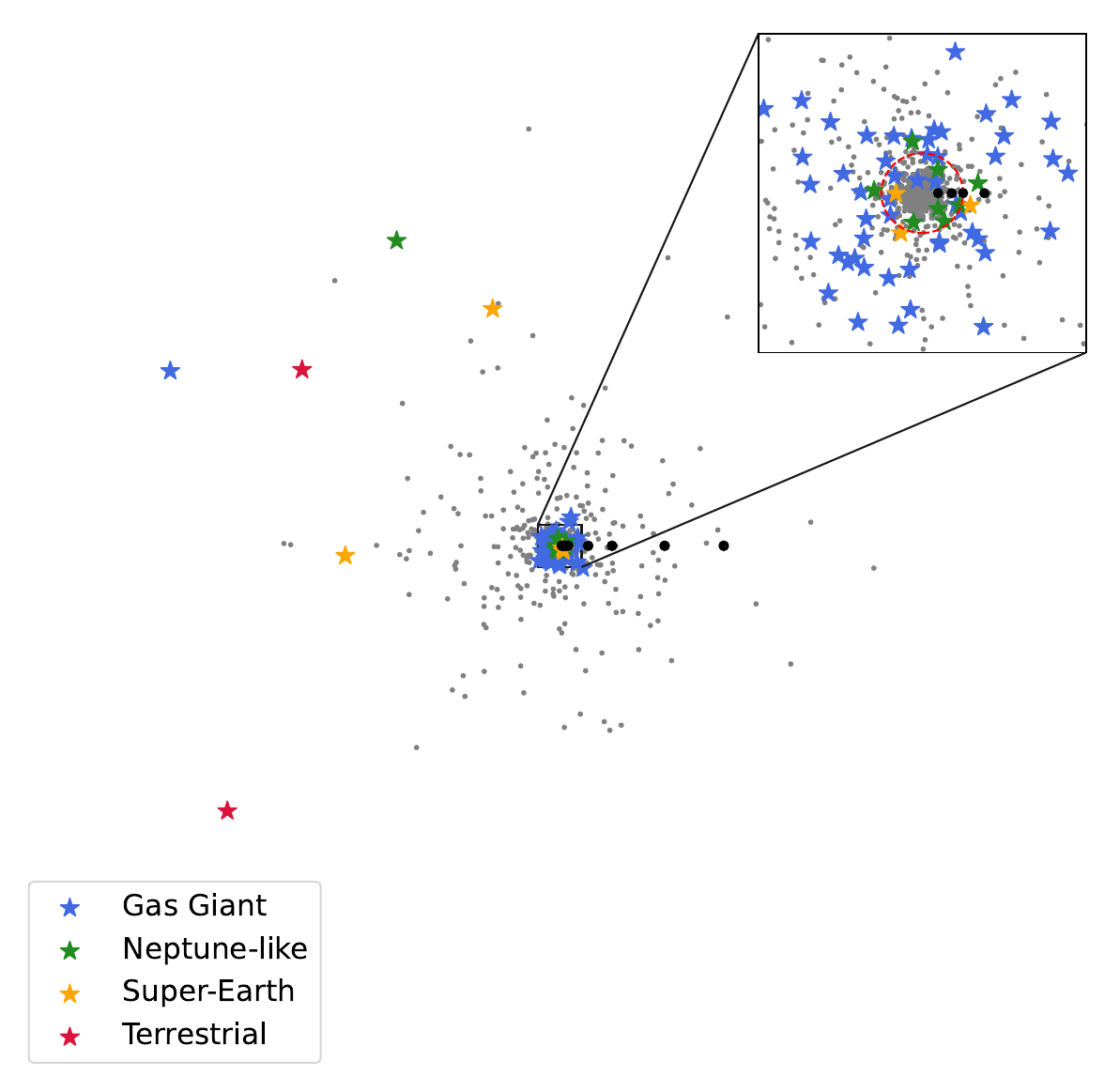}
    \label{rgb.fig}}
    \caption{Left panel: The scaled distances between planets and their host stars in current state (e.g., main sequence or giant). The grey points represent all planets. The blue, green, orange, and red asterisks represent gas giants, Neptunian-like planets, super-Earths, and terrestrial planets that are located within both the CHZ and UHZ. The distances are scaled by dividing by the stellar radii. All stellar radii are normalized to 1 solar radius, represented by the dashed red circle, for better visual effect. Right panel: The scaled distances between planets and their host stars at the end of the red giant phase.}
    \label{evol.fig}
\end{figure*}

Figure \ref{ms.fig} shows the scaled distances between the planets and their host stars at current stage, while Figure \ref{rgb.fig} shows the scaled distances at the end of the red giant phase for exoplanets orbiting stars that can evolve to the red giant branch within the age of the universe.
Here the distances were scaled by dividing by the stellar radii, with current stellar radii taken from the PSCD catalog and those at the end of the red giant phase derived from the MIST models.
For better visual visualization, all stellar radii were normalized to 1 solar radius (i.e., the dashed red circle).
We found that the majority of the planets, including Earth, would be engulfed by their host stars during the red giant phase. Only about 15\% of the planets are likely to survive. 

Interestingly, it was found that most planets located within both the CHZ and the UHZ would not be engulfed. This suggests that planets in the overlapping region of the CHZ and UHZ are more likely to survive the later stages of stellar evolution, possibly due to their positions being farther from their host stars compared to other planets. 
These planets are likely to maintain stable orbits and potentially retain conditions suitable for habitability as their host stars evolve into the red giant phase.

\section{Stellar Activity of Host Stars}
\label{activity.sec}

The UV activity indices are also important for understanding the radiation environment of exoplanets and assessing their potential habitability. 
We used the same method as described in \cite{2024ApJ...966...69L} to calculate stellar UV activity index following 
\begin{equation}
\label{activity_index.eq}
    R^{\prime}_{\rm UV} = \frac{f_{\rm UV,exc}}{f_{\rm bol}} = \frac{f_{\rm UV,obs}\times(\frac{d}{R})^2-f_{\rm UV,ph}}{f_{\rm bol}},
\end{equation}
where $f_{\rm UV,exc}$ represents the UV excess flux due to magnetic activity, $f_{\rm bol}$ is the bolometric flux determined using the effective temperature with $f_{\rm bol} = \sigma_{B} T_{\rm eff}^{4}$, $f_{\rm UV,obs}$ is the extinction-corrected flux derived from the observed FUV or NUV magnitudes and corresponding extinction values, $f_{\rm UV,ph}$ refers to the photospheric flux at stellar surface, and $R$ is stellar radius obtained from the PSCD catalog. We used the method presented by \cite{wang2024predictingphotosphericuvemission} to estimate the photospheric flux ($f_{\rm UV,ph}$), which provides a more convenient and efficient formula for estimating stellar UV photospheric flux for F to M-type main-sequence stars. 
They used PARSEC \citep[PAdova and TRieste Stellar Evolution Code;][]{2012MNRAS.427..127B} models to establish relations between NUV and FUV magnitudes and stellar parameters, such as effective temperature and $Gaia$ BP$-$RP color, for dwarfs with different metallicities. These relationships are described by tenth-order polynomials\footnote{\url{https://github.com/AstroSong/UVphotosphere}}.

\begin{figure*}[htp]
    \centering
    \subfigure[]{
    \includegraphics[width=0.48\textwidth]{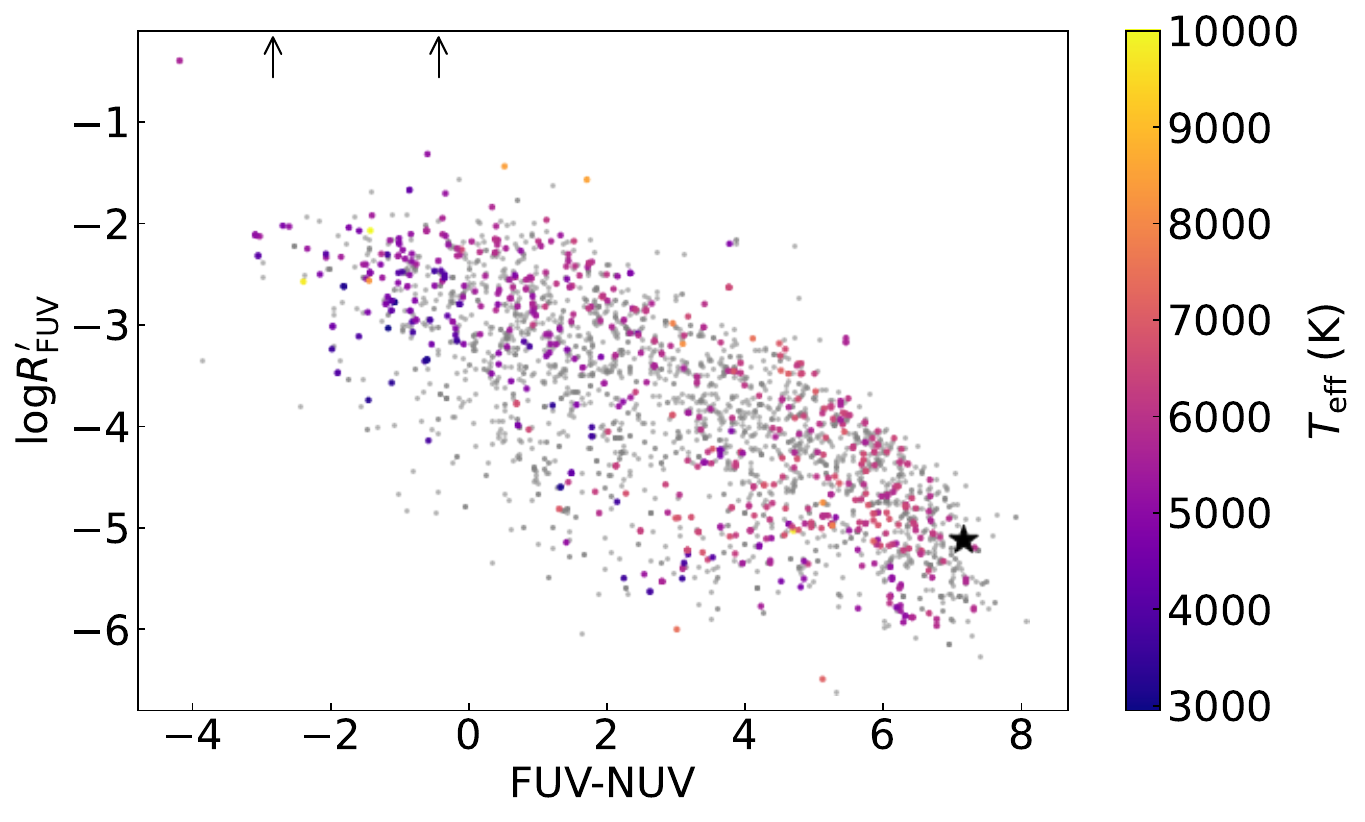}
    \label{fuv_R.fig}}
    \subfigure[]{
    \includegraphics[width=0.48\textwidth]{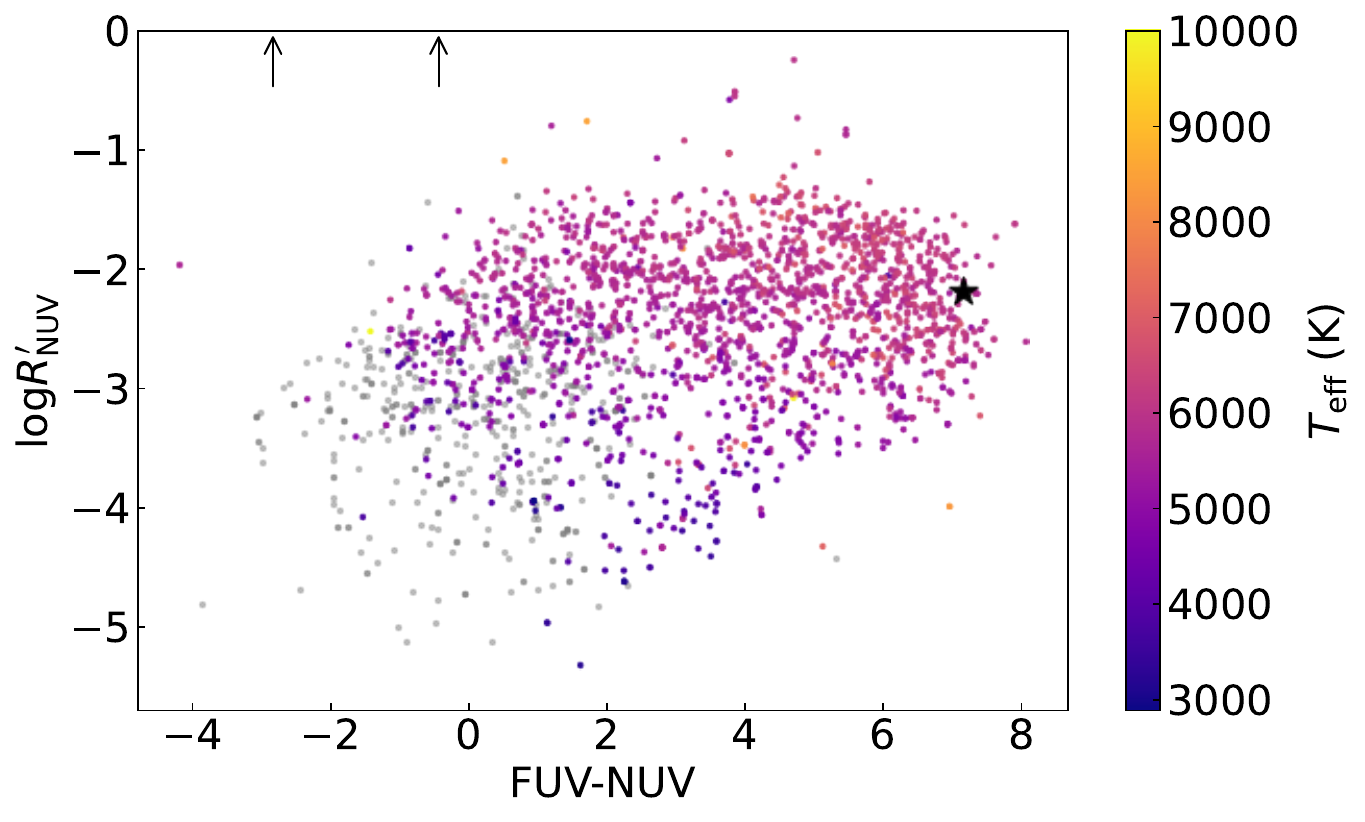}
    \label{nuv_R.fig}}
    \caption{Left panel: The relationship between the activity index $R^{\prime}_{\rm FUV}$ and the FUV$-$NUV color. Gray points represent saturated sources and sources with upper limit magnitudes, while colored points represent sources with general photometry. The colorbar represents effective temperature of stars. The black asterisk is the Sun. Upward arrows mark the UV emission of two hot subdwarfs exceeding the plotted range. Right panel: The relationship between the activity index $R^{\prime}_{\rm NUV}$ and the FUV$-$NUV color.}
    \label{activity.fig}
\end{figure*}

Figures \ref{fuv_R.fig} and \ref{nuv_R.fig} show the relationships between FUV and NUV activity indices and the FUV$-$NUV color, respectively. The FUV-NUV index is also often used as an indicator of stellar activity \citep{2006A&A...458..921W,2016ApJ...820...89F,2018ApJS..235...16B}. 
There is a clear negative correlation between $R^{\prime}_{\rm FUV}$ and FUV$-$NUV.
This is because FUV emission primarily originates from the chromosphere and transition region and is dominated by magnetic activity.
However, no clear relationship is found between $R^{\prime}_{\rm NUV}$ and FUV$-$NUV. 
This suggests that chromospheric NUV emission is not only related to magnetic activity, but may also be influenced by non-magnetic heating mechanisms, such as acoustic waves \citep{1996SSRv...75..453N,1997ApJ...481..500C}.
Therefore, $R^{\prime}_{\rm FUV}$ and FUV$-$NUV are more reliable indicators of stellar magnetic activity.



\begin{figure*}[htp]
    \centering
    \subfigure[]{
    \includegraphics[width=0.45\textwidth]{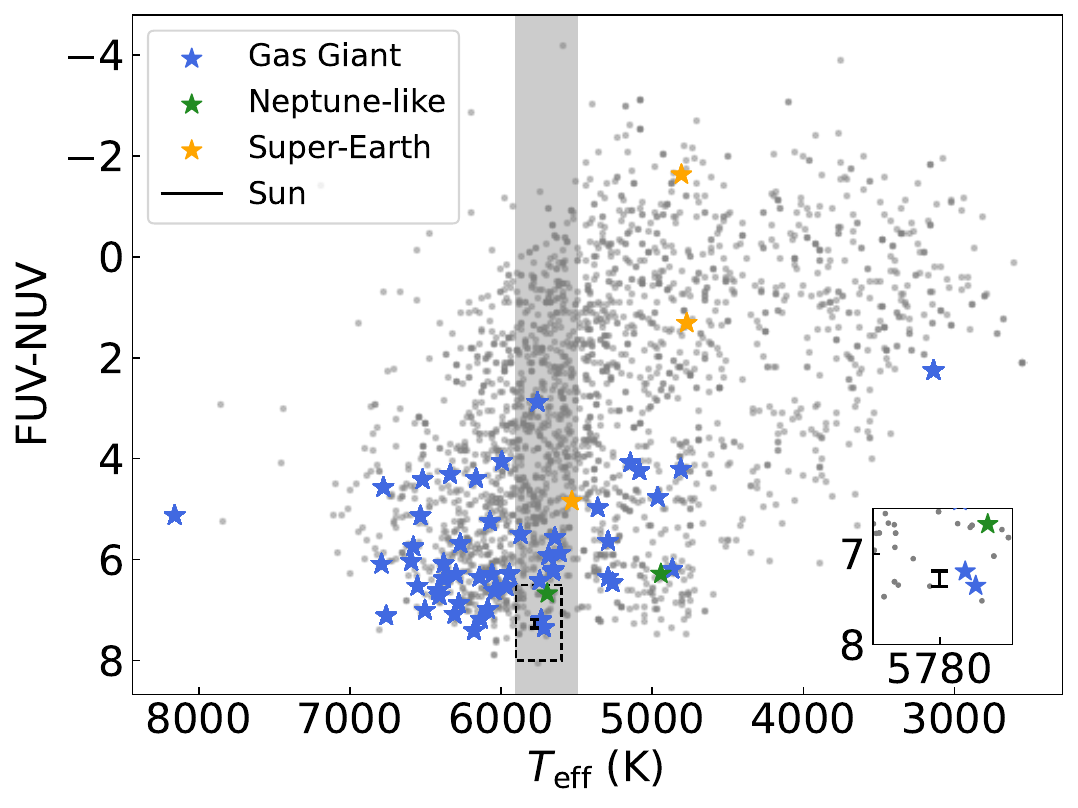}
    \label{hr_hz.fig}}
    \subfigure[]{
    \includegraphics[width=0.45\textwidth]{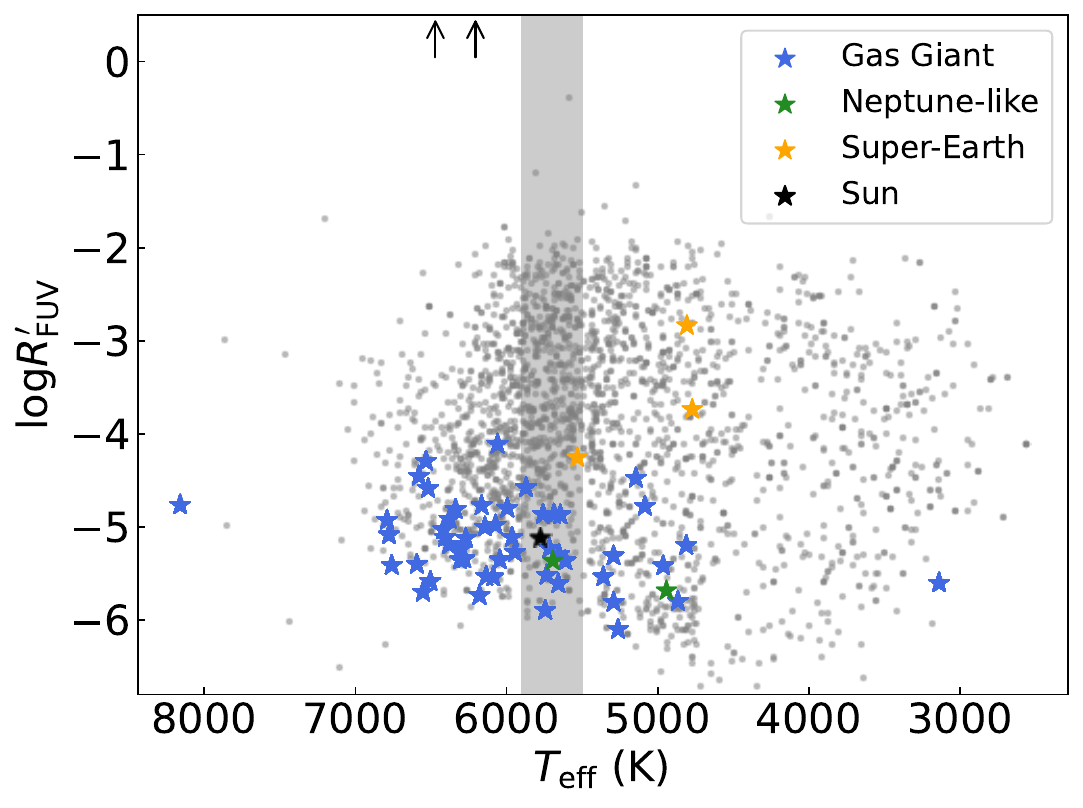}
    \label{r_teff_fuv.fig}}
    \caption{Left panel: FUV$-$NUV color versus the effective temperature of the sample stars. 
    The blue, green, and orange points represent stars with gas giants, Neptunian-like planets, and super-Earths that are located within both the CHZ and UHZ, respectively. The black bar represents the FUV$-$NUV variation of the Sun over one activity cycle \citep{2017JSWSC...7A...6L}.} The shaded region marks the temperate range from 5500 to 5900 K, representing solar-like stars. Right Panel: $R^{\prime}_{\rm FUV}$ versus the effective temperature of the sample stars. The black star is the Sun.
    \label{r_teff.fig}
\end{figure*}

Figure \ref{hr_hz.fig} shows the relationships between FUV$-$NUV, $R^{\prime}_{\rm FUV}$, and the temperatures of host stars with confirmed exoplanets.
It's clear that K and M stars have much larger FUV$-$NUV values than F and G stars, suggesting higher activity of cooler stars.
However, the relation between $R^{\prime}_{\rm FUV}$ and $T_{\rm eff}$ is not clear.
We concluded that among the UV indices, FUV$-$NUV is a more effective measure of stellar activity than $R^{\prime}_{\rm FUV}$ and $R^{\prime}_{\rm NUV}$.
Notably, the Sun exhibits a quite large FUV$-$NUV value and a low $R^{\prime}_{\rm FUV}$ value, even among solar-like stars (i.e., 5500 K $<$ $T_{\rm eff}$ $<$ 5900 K).
Furthermore, the variation of FUV$-$MUV for the Sun during one activity cycle is quite small, ranging from approximately 7.188 to 7.355 \citep{2017JSWSC...7A...6L}.
This suggests that our Sun has very low FUV emission, indicative of low magnetic activity, posing minimal risk to human beings. 
Simultaneously, the moderate NUV emission level of the Sun, as shown in Figure \ref{nuv_R.fig}, contributes to conditions that are favorable for life on Earth.

\section{Summary}
\label{sum.sec}

In this study, we conducted a detailed analysis of the UV photometry of host stars with confirmed planets, using observational data from the GALEX and Swift UVOT missions. 

For both GALEX and SWIFT data, we performed aperture photometry on single-exposure images using \texttt{photutils}. For sources identified as saturated through radial profile analysis, we applied a PSF fitting technique to calculate the total counts using PSF models provided by GALEX. For sources close to detection limits, the upper limit magnitudes were given.
Finally, we obtained UV photometry for 2742 host stars with the GALEX observations and 27 host stars with the SWIFT UVOT observations.
The catalogs can be used to explore a wide range of scientific questions, such as stellar UV activity and planet habitability.

We calculated the CHZ and UHZ for our sample stars. 
Most of the detected exoplanets do not reside within the CHZ and UHZ but are located closer to their host stars. Only 86 exoplanets were found within both the CHZ and UHZ, and the majority of these are gas giants. 
Additionally, as the host stars evolve into the red giant phase, most planets, including Earth, are likely to be engulfed. However, planets that lie within both the CHZ and UHZ are less likely to be swallowed, suggesting they may potentially maintain stable orbits and retain conditions suitable for habitability over a long timescale. 

We also calculated the stellar UV activity of host stars with GALEX observations. There is a linear but diffuse relationship between the FUV activity index and the FUV$-$NUV color. FUV$-$NUV is more effective at representing stellar activity than $R^{\prime}_{\rm FUV}$ and $R^{\prime}_{\rm NUV}$. 
The Sun's low FUV emission and moderate NUV emission suggest it is unique among stars, even compared to other solar-like stars. 

We cross-matched our stellar sample with the NASA Exoplanet Exploration Program (ExEP) Target List for the Habitable Worlds Observatory (HWO) \citep{2024arXiv240212414M}. In our sample, we identified 22 stars that are also listed as high-priority targets for HWO observations, which includes 7 priority A targets (47 UMa, HD 102365, HD 141004, HD 192310, HD 219134, HD 26965, eps Ind A), 9 priority B targets (61 Vir, GJ 411, GJ 887, HD 160691, HD 20794, HD 3651, HD 39091, HR 810, rho CrB), and 6 priority C targets (55 Cnc, HD 140901, HD 189567, HD 69830, HN Peg, eps Eri). These stars are of particular interest for further study.

\section*{acknowledgements}

We thank the anonymous referee for helpful comments and suggestions that have significantly improved the paper. 
The Guoshoujing Telescope (the Large Sky Area Multi-Object Fiber Spectroscopic Telescope LAMOST) is a National Major Scientific Project built by the Chinese Academy of Sciences. Funding for the project has been provided by the National Development and Reform Commission. LAMOST is operated and managed by the National Astronomical Observatories, Chinese Academy of Sciences.
GALEX data presented in this paper were obtained from the Mikulski Archive for Space Telescopes (MAST).
This work presents results from the European Space Agency (ESA) space mission {\it Gaia}. {\it Gaia} data are being processed by the {\it Gaia} Data Processing and Analysis Consortium (DPAC). Funding for the DPAC is provided by national institutions, in particular the institutions participating in the {\it Gaia} MultiLateral Agreement (MLA). We acknowledge use of the VizieR access tool, operated at CDS, Strasbourg, France, and of Astropy, a community-developed core Python package for Astronomy (Astropy Collaboration, 2013). 
This work was supported by National Key Research and Development Program of China (NKRDPC) under grant Nos. 2019YFA0405000, Strategic Priority Program of the Chinese Academy of Sciences undergrant No. XDB4100000, Science Research Grants from the China Manned Space Project with No. CMS-CSST-2021-A08, and National  Natural Science Foundation of China (NSFC) under grant Nos. 11988101/11933004/11833002/12090042/12273057. S.W. acknowledges support from the Youth Innovation Promotion Association of the CAS (IDs 2019057).

\bibliographystyle{yahapj}
\bibliography{bibtex.bib}{}

\begin{appendix}

\renewcommand{\thetable}{A\arabic{table}}
\setcounter{table}{0}

\section{PSF Model Fitting}
\label{fit.sec}

We did a fifteenth-order polynomial fitting of the radial profiles of the PSF models in both FUV and NUV bands, following 
\begin{equation*}
    f(x) = c_{15} \cdot x^{15} + c_{14} \cdot x^{14} + \cdots + c_1 \cdot x + c_0,
\end{equation*}
where \( x \) is the radius in pixels, \( f(x) \) is the relative flux, with units of erg/s per pixel, and \( c_0, c_1, \dots, c_{15} \) are the coefficients obtained from the polynomial fitting process. The fitting coefficients for the FUV and NUV bands are listed in Table \ref{psf.tab}.

\begin{table*}[htp]
\centering
\caption{The fifteenth-order polynomial fitting coefficients of the PSF models.}
\begin{tabular}{*{3}{c}}
\hline\noalign{\smallskip}
Coefficients & FUV & NUV\\
\hline\noalign{\smallskip}
$ c _{ 15 } $ & -3.5193074265e-18 & -6.5368646814e-18 \\
$ c _{ 14 } $ & 1.0347819972e-15 & 1.8424745388e-15 \\
$ c _{ 13 } $ & -1.3904411453e-13 & -2.3364149832e-13 \\
$ c _{ 12 } $ & 1.1334605585e-11 & 1.7648404347e-11 \\
$ c _{ 11 } $ & -6.2722876813e-10 & -8.8708667366e-10 \\
$ c _{ 10 } $ & 2.4951484069e-08 & 3.1439641284e-08 \\
$ c _{ 9 } $ & -7.3563228459e-07 & -8.1363384886e-07 \\
$ c _{ 8 } $ & 1.6286799964e-05 & 1.5731825277e-05 \\
$ c _{ 7 } $ & -2.7050738122e-04 & -2.3044706481e-04 \\
$ c _{ 6 } $ & 3.3199046435e-03 & 2.5618904757e-03 \\
$ c _{ 5 } $ & -2.9160403940e-02 & -2.1197794071e-02 \\
$ c _{ 4 } $ & 1.7321509392e-01 & 1.2365331725e-01 \\
$ c _{ 3 } $ & -6.2884886458e-01 & -4.5558846047e-01 \\
$ c _{ 2 } $ & 1.1521859086e+00 & 8.6017736078e-01 \\
$ c _{ 1 } $ & -1.0293797186e+00 & -8.1743725644e-01 \\
$ c _{ 0 } $ & 1.5680723718e-02 & 2.2836177105e-02 \\
\hline\noalign{\smallskip}
\end{tabular}
\label{psf.tab}
\end{table*}

\clearpage

\section{Comparison of Upper Limits}
\label{upperlimit_appendix.sec}
\renewcommand{\thefigure}{B\arabic{figure}}
\setcounter{figure}{0}

\begin{figure*}[htp]
    \centering 
    \subfigure[FUV band]{
    \includegraphics[width=0.46\textwidth]{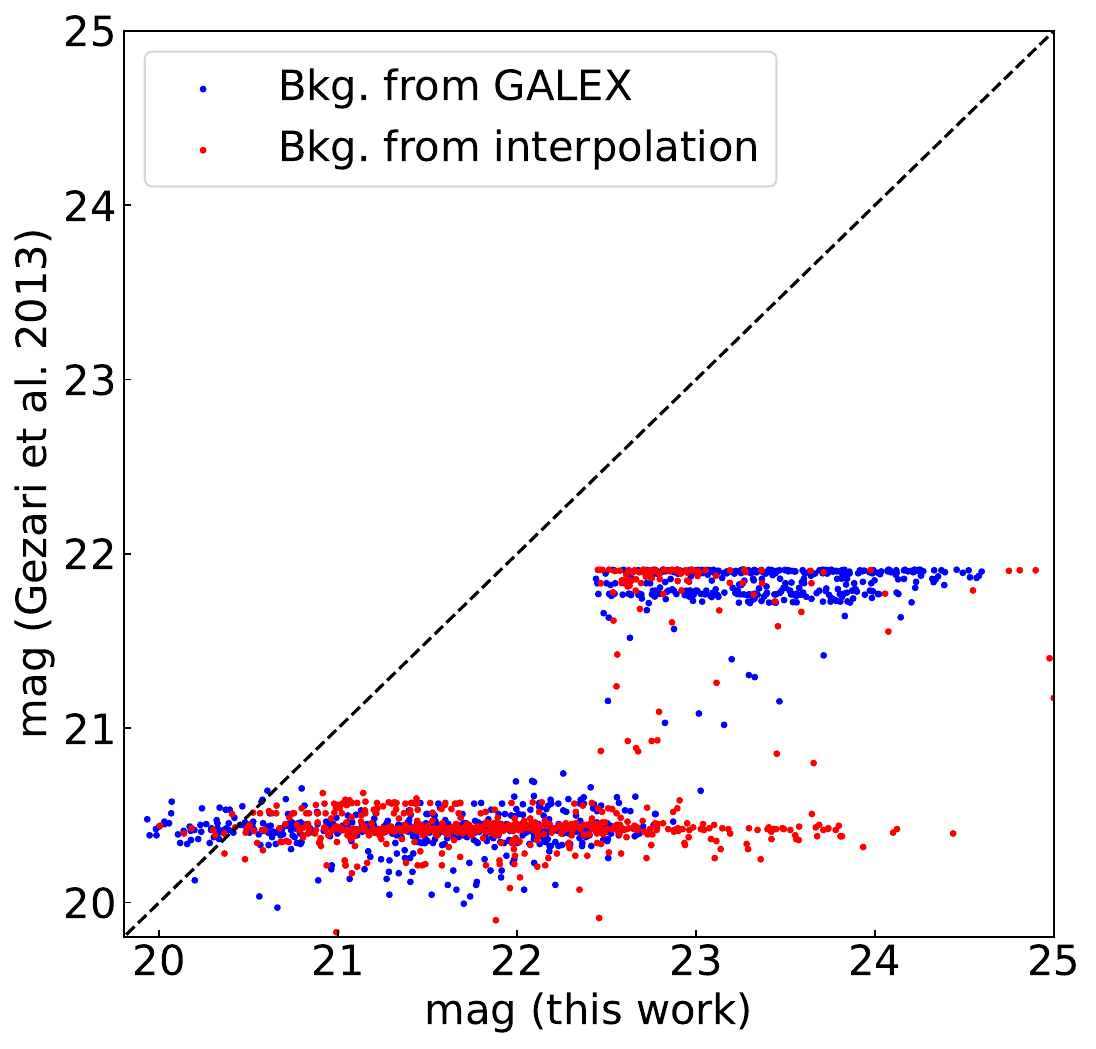}
    \label{fuv_upperlimit.fig}}
    \subfigure[NUV band]{
    \includegraphics[width=0.46\textwidth]{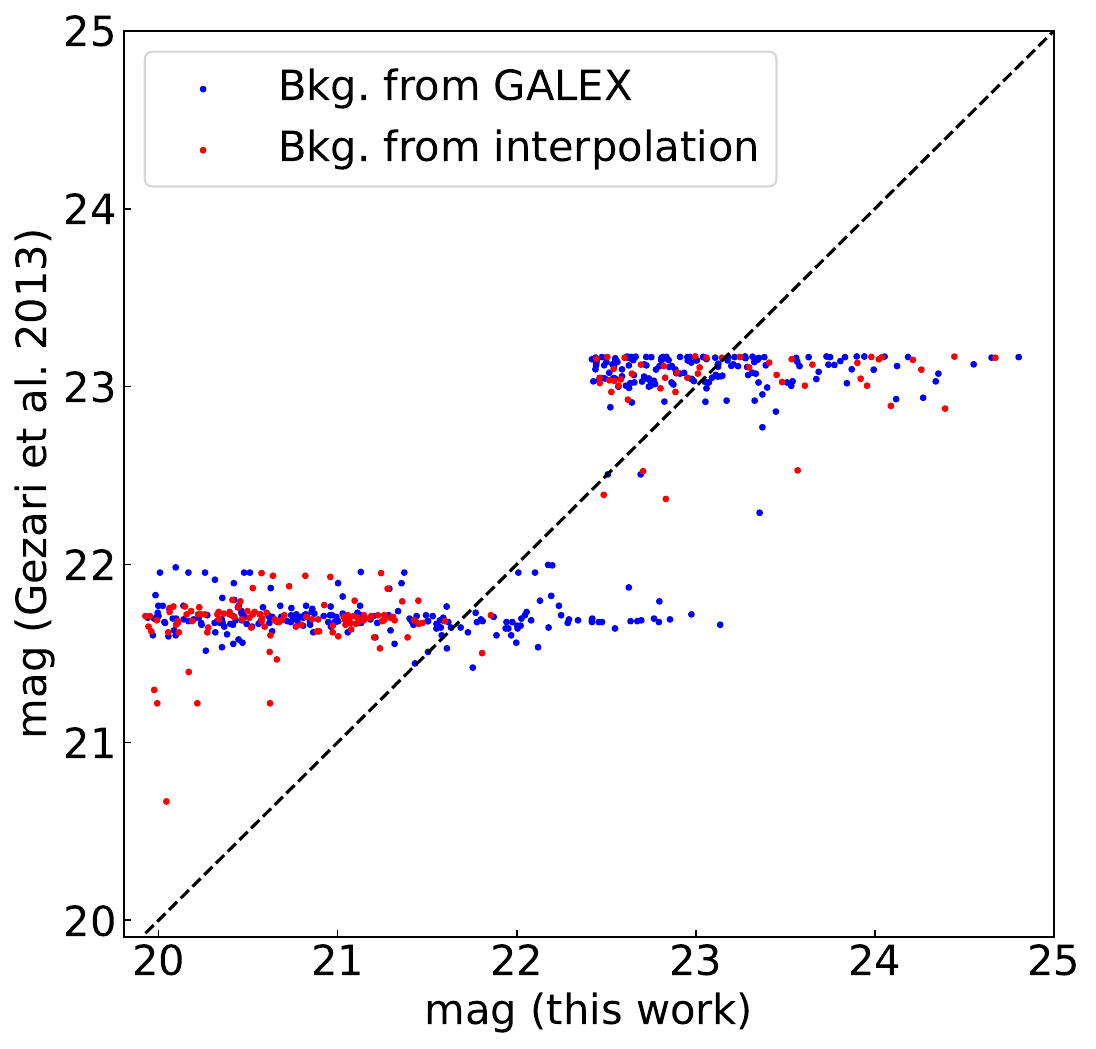}
    \label{nuv_upperlimit.fig}}
    \subfigure[FUV band]{
    \includegraphics[width=0.46\textwidth]{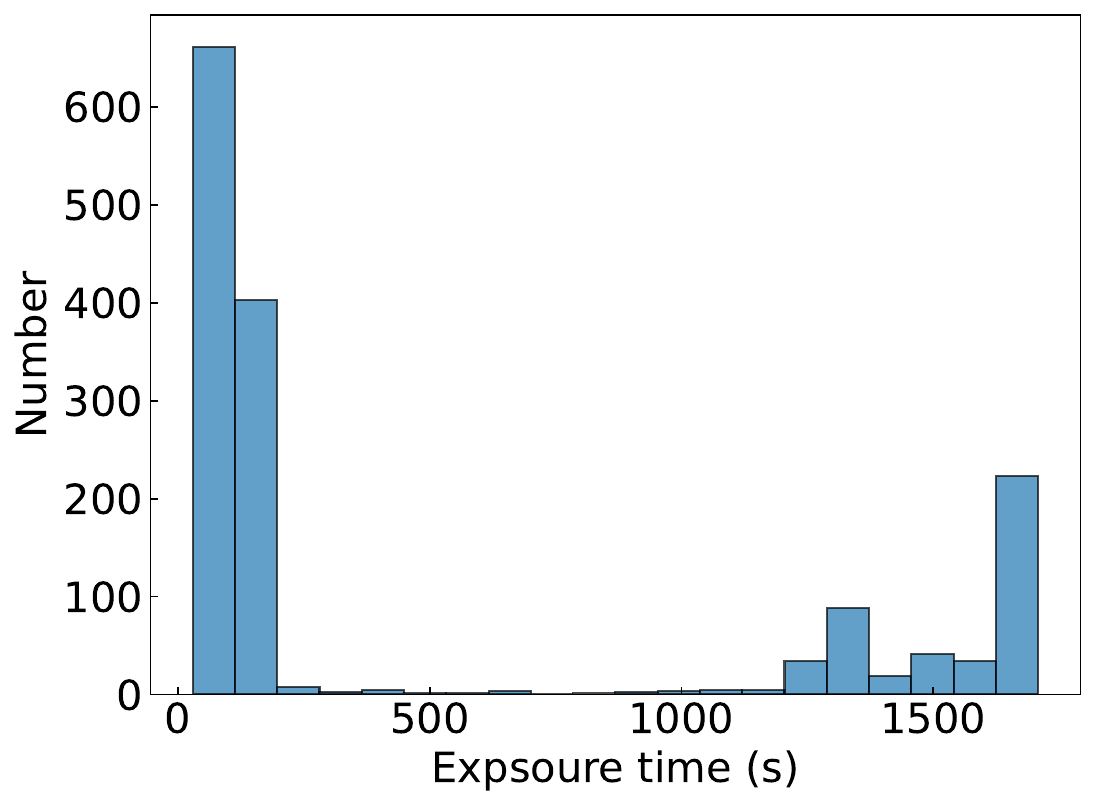}
    \label{fuv_upperlimit_exptime_hist.fig}}
    \subfigure[NUV band]{
    \includegraphics[width=0.46\textwidth]{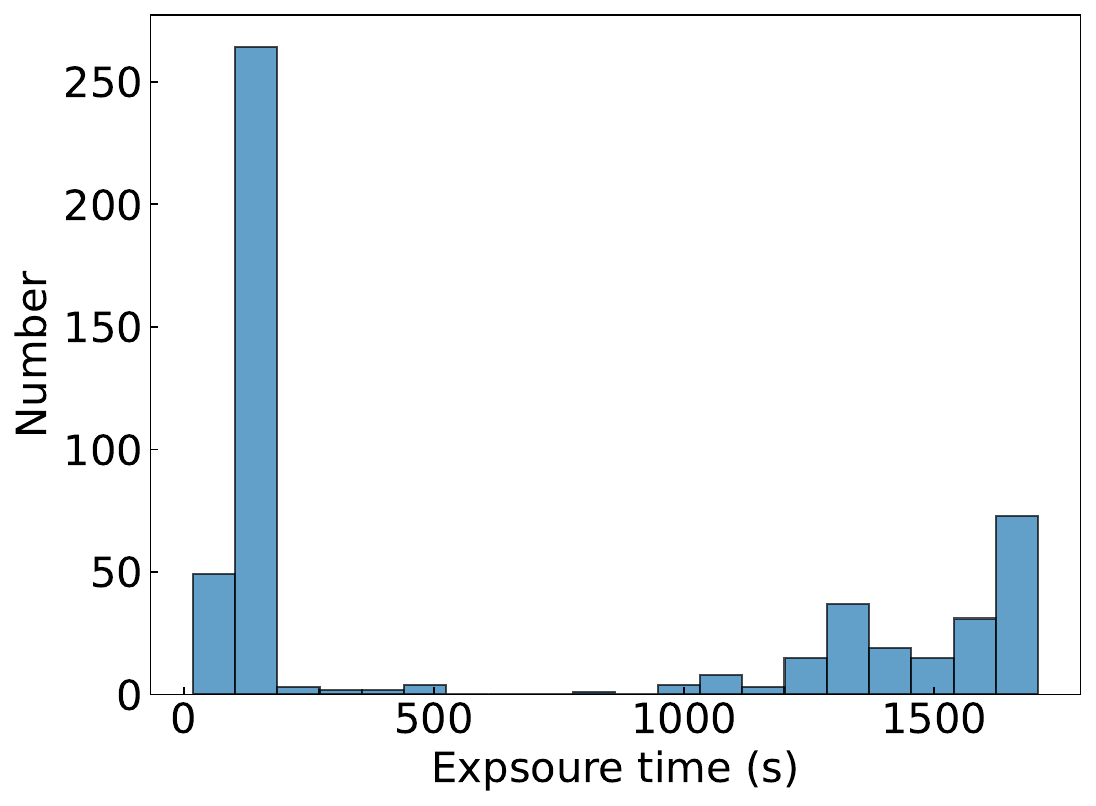}
    \label{nuv_upperlimit_exptime_hist.fig}}
    \caption{Top panel: Comparison between the upper limit magnitudes from our aperture photometry with those calculated according to \cite{2013ApJ...766...60G} in the FUV (Left) and NUV (Right) bands. Blue points represent the background values obtained from the GALEX archive, and red points represent background estimates with the interpolation method. 
    Bottom panel: Distribution of exposure times in the FUV (Left) and NUV (Right) bands.}
    \label{upperlimit.fig}
\end{figure*}

Figure \ref{upperlimit.fig} compares the upper limit magnitudes we calculated with those calculated using the formula from \cite{2013ApJ...766...60G}.

\clearpage

\section{Comparison of Atmospheric Parameters}
\label{pars_sub.sec}

\renewcommand{\thefigure}{C\arabic{figure}}
\setcounter{figure}{0}

\begin{figure*}[htp]
    \centering
    \subfigure[]{
    \includegraphics[width=0.31\textwidth]{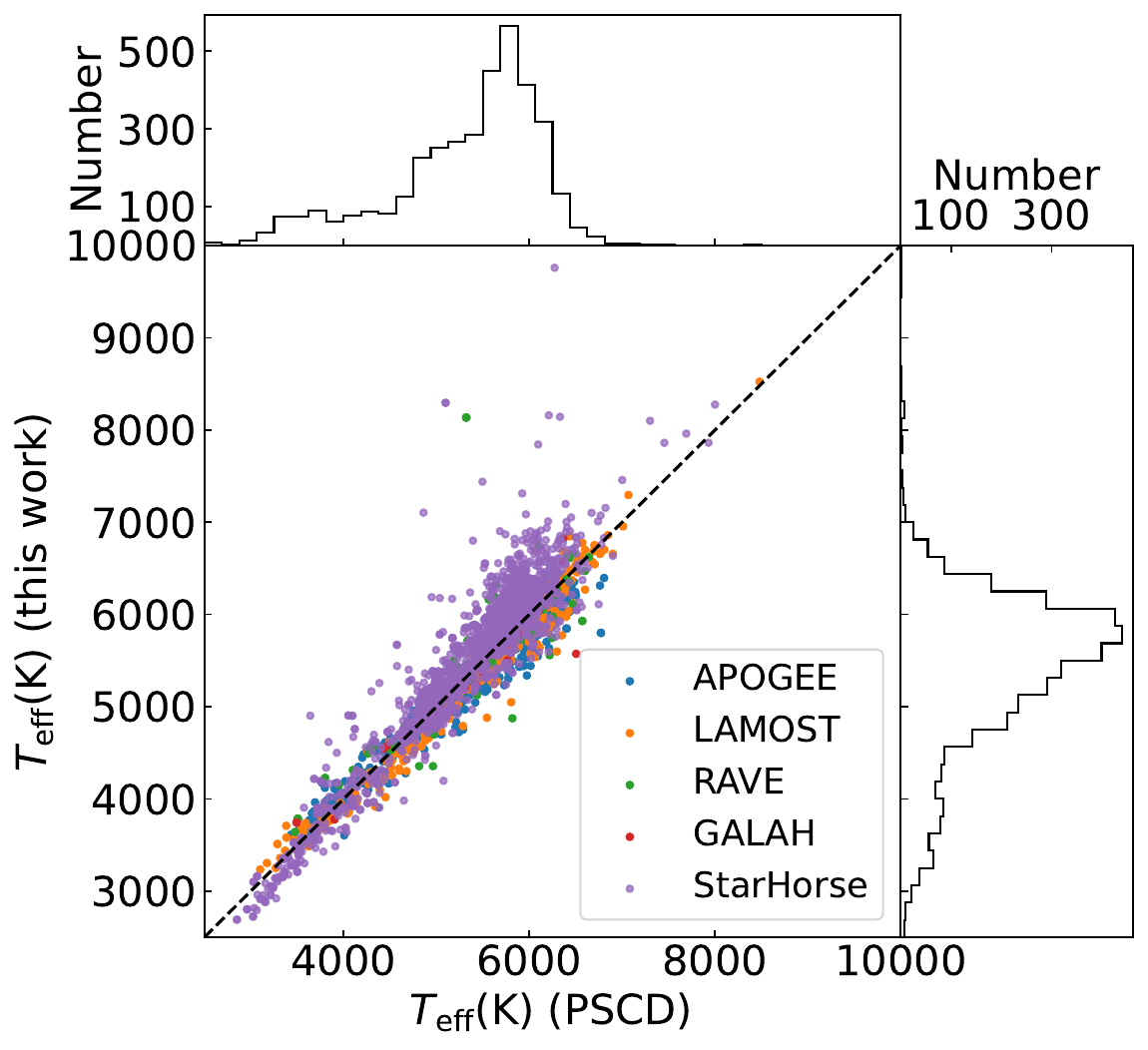}
    \label{teff.fig}}
    \subfigure[]{
    \includegraphics[width=0.31\textwidth]{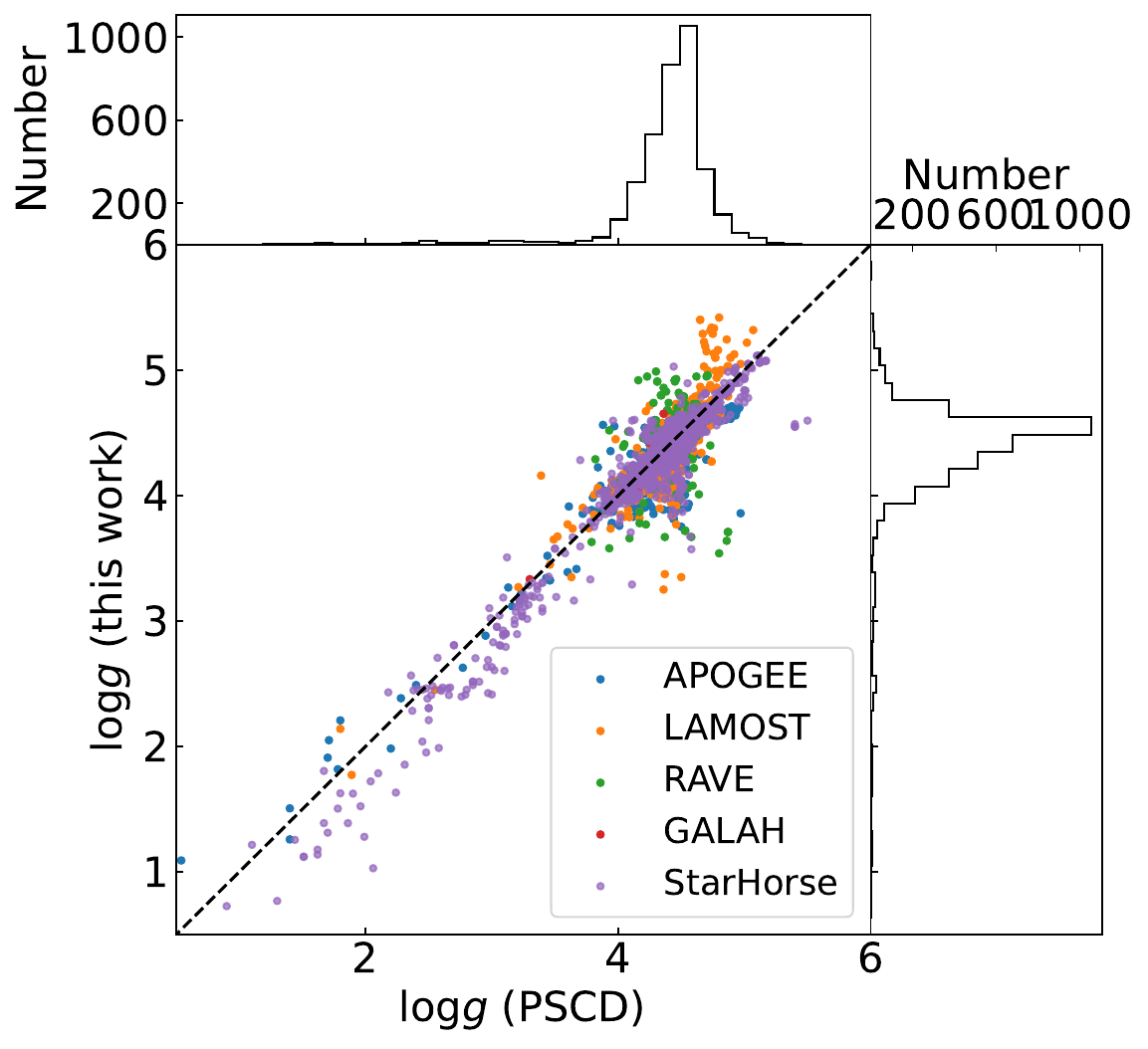}
    \label{logg.fig}}
    \subfigure[]{
    \includegraphics[width=0.31\textwidth]{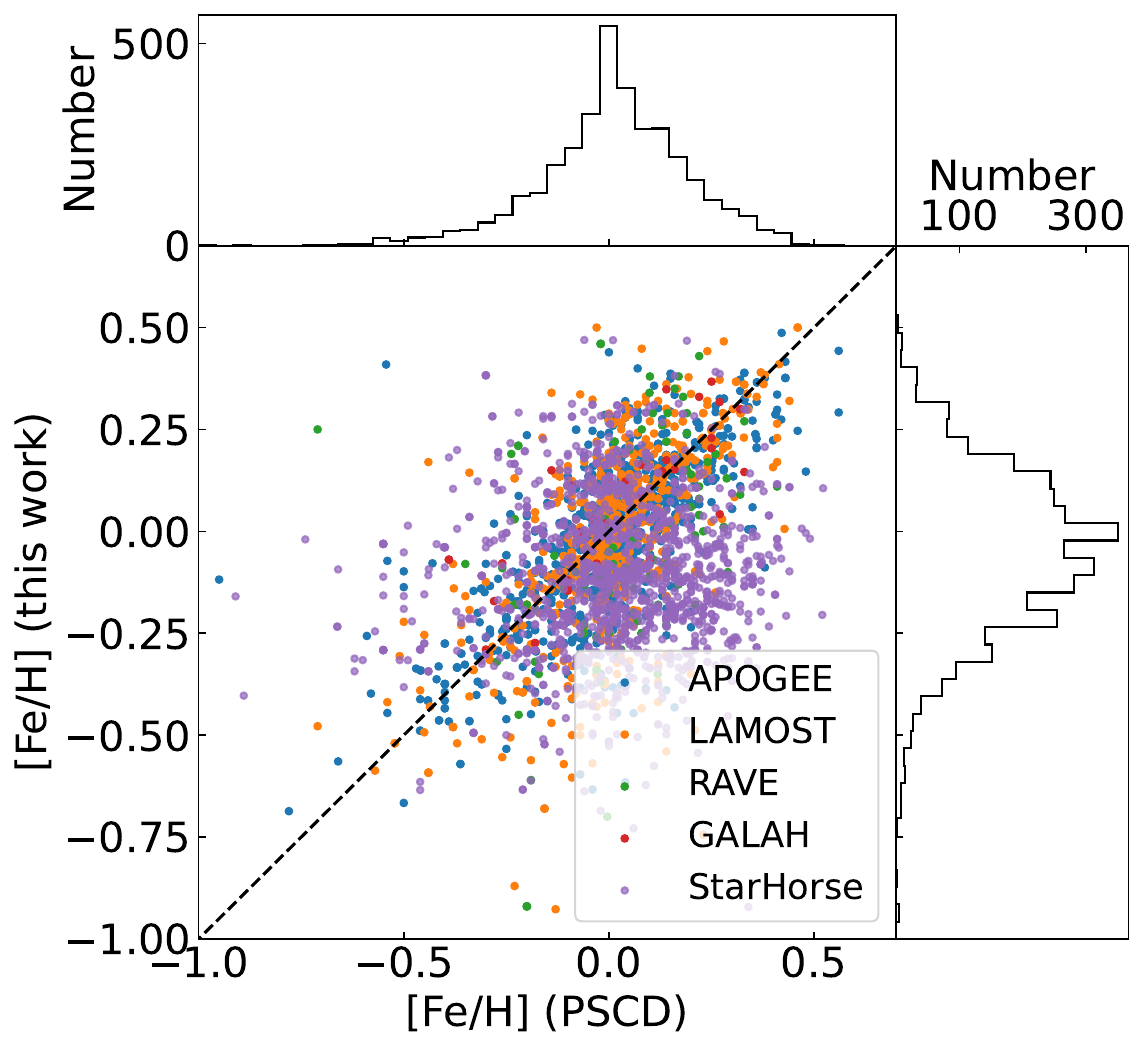}
    \label{feh.fig}}
    \caption{Comparison of atmospheric parameters from different catalogs and those from \cite{ps}. The panels from left to right display effective temperature, surface gravity, and metallicity, respectively. Above the main plot is the distribution of the parameters from \cite{ps}. To the right of main plot is the distribution of the parameters used in this paper.}
    \label{pars.fig}
\end{figure*}

Figure \ref{pars.fig} compares the atmospheric parameters from various stellar catalogs with those provided by the PCSD catalog from \cite{ps}. 

\clearpage

\section{Flares in the sample}
\label{flare.sec}
\renewcommand{\thetable}{D\arabic{table}}
\setcounter{table}{0}

We also calculated the energy of each flare event using the following equation \citep{2023MNRAS.519.3564J}:

\begin{equation}
    E = 4 \pi d^{2} \delta \lambda \int_{t_{\rm start}}^{t_{\rm end}} \left( F(t) - F_{0} \right) dt,
\end{equation}

where $d$ represents the distance from Gaia eDR3, and $\delta \lambda$ corresponds to the effective bandwidths of the FUV and NUV bands. The values $t_{\rm start}$ and $t_{\rm end}$ indicate the times when the flare begins and ends, respectively. $F(t)$ is the observed flux at each time point during the flare, while $F_{0}$ is the quiescent flux. The quiescent flux was determined iteratively by removing outliers and calculating the median of the remaining points.

\begin{table*}[htp]

\begin{center}
\caption{The flare events in the sample, including FUV and NUV bands.}
\label{flare.tab}
\begin{tabular}{lcccccc}
\hline\noalign{\smallskip}
Host name & $t_{\rm start}$ & $\rm t_{\rm end}-t_{\rm start}$ & Peak luminosity & Flare energy & Equivalent Duration & Band \\
 & (s) & (s)  & log (erg/s) & log (erg) & (s) & \\
\hline
\multirow{4}*{24 Sex} & 920501083.995 & 100 & 31.58 & 33.56 & 102 & NUV \\
 & 949723669.995 & 200 & 31.57 & 33.86 & 205 & NUV \\
 & 922393771.995 & 160 & 31.60 & 33.77 & 167 & NUV \\
 & 953089745.995 & 160 & 31.59 & 33.77 & 165 & NUV \\
 \hline\noalign{\smallskip}
2MASS J12073346-3932539 & 989007303.995 & 80 & 29.24 & 30.94 & 182 & NUV \\
\hline\noalign{\smallskip}
\multirow{2}*{BD+14 4559} & 772780784.995 & 140 & 29.89 & 31.96 & 147 & NUV \\
  & 772774728.995 & 200 & 29.89 & 32.13 & 219 & NUV \\
\hline\noalign{\smallskip}
GJ 273 & 883514153.995 & 100 & 26.84 & 28.73 & 121 & NUV \\
Gl 378 & 855139897.995 & 100 & 28.25 & 30.13 & 139 & NUV \\
HAT-P-16 & 780278683.995 & 120 & 30.12 & 31.72 & 248 & FUV \\
HAT-P-18 & 934078056.995 & 60 & 30.41 & 31.94 & 92 & NUV \\
HD 155918 & 827072563.995 & 100 & 29.09 & 30.83 & 136 & FUV \\
HD 156279 & 899806945.995 & 120 & 28.96 & 30.62 & 661 & FUV \\
HD 158259 & 933574033.995 & 320 & 30.50 & 32.90 & 493 & NUV \\
HD 216520 & 1000529790.995 & 140 & 30.14 & 32.27 & 143 & NUV \\
HD 217786 & 907539355.995 & 120 & 29.34 & 31.21 & 205 & FUV \\
\hline\noalign{\smallskip}
\multirow{2}*{HD 3765} & 780290639.995 & 120 & 28.05 & 29.80 & 254 & FUV \\
  & 934988026.995 & 140 & 30.07 & 32.13 & 149 & NUV \\
\hline\noalign{\smallskip}
HD 5583 & 846957175.995 & 100 & 32.02 & 33.97 & 105 & NUV \\
HD 6860 & 904758566.995 & 140 & 30.03 & 31.83 & 482 & FUV \\
\hline\noalign{\smallskip}
\end{tabular}
\end{center}
    NOTE. This table is available in its entirety in machine-readable and VO forms in the online journal. A portion is shown here for guidance regarding its form and content.

\end{table*}

\end{appendix}

\end{document}